\documentclass[aps,prl,preprint,groupedaddress]{revtex4-1}
\usepackage{graphicx}
\usepackage{color}
\usepackage{xfrac}
\usepackage{amsmath,bm}
\usepackage{lineno}

\begin{document}

\title{Coherent generation of symmetry-forbidden phonons by light-induced electron-phonon interactions in magnetite}

\author{S. Borroni$^{1}$, E. Baldini$^{1,2}$, V. M. Katukuri$^{3}$, A. Mann$^{1}$, K. Parlinski$^{4}$, D. Legut$^{5}$, C. Arrell$^{2}$, F. van Mourik$^{2}$, J. Teyssier$^{6}$, A. Kozlowski$^{7}$, P.~Piekarz$^{4}$, O. V. Yazyev$^{3}$, A. M. Ole\'{s}$^{8}$, J. Lorenzana$^{9}$, F. Carbone$^{1,*}$}

\affiliation{$^{1}$Laboratory for Ultrafast Microscopy and Electron Scattering and the Lausanne Centre for Ultrafast Science, IPHYS, \'{E}cole Polytechnique F\'{e}d\'{e}rale de Lausanne, CH-1015 Lausanne, Switzerland}
\affiliation{$^{2}$Laboratory for Ultrafast Spectroscopy and the Lausanne Centre for Ultrafast Science, ISIC, \'{E}cole Polytechnique F\'{e}d\'{e}rale de Lausanne, CH-1015 Lausanne, Switzerland}
\affiliation{$^{3}$Chair of Computational Condensed Matter Physics, IPHYS, \'{E}cole Polytechnique F\'{e}d\'{e}rale de Lausanne, CH-1015 Lausanne, Switzerland}
\affiliation{$^{4}$Institute of Nuclear Physics, Polish Academy of Sciences, 31-342 Krak\'{o}w, Poland}
\affiliation{$^{5}$IT4Innovations Center, VSB-Technical University of Ostrava, 17.listopadu 15, 708 33 Ostrava, Czech Republic}
\affiliation{$^{6}$Department of Quantum Matter Physics, University of Geneva, CH-1211 Geneva, Switzerland}
\affiliation{$^{7}$Faculty of Physics and Applied Computer Science, AGH-University of Science and Technology, Aleja Mickiewicza 30, PL-30059 Krak\'{o}w, Poland}
\affiliation{$^{8}$Marian Smoluchowski Institute of Physics, Jagiellonian University, prof. S. Lojasiewicza 11, PL-30348 Krak\'{o}w, Poland}
\affiliation{$^{9}$Institute for Complex Systems - CNR, and Physics Department, University of Rome ``La Sapienza'', I-00185 Rome, Italy}

\begin{abstract}
Symmetry breaking across phase transitions often causes changes in selection rules and emergence of optical modes which can be detected via spectroscopic techniques or generated coherently in pump-probe experiments. In second-order or weakly first-order transitions, fluctuations of the order parameter are present above the ordering temperature, giving rise to intriguing precursor phenomena, such as critical opalescence. Here, we demonstrate that in magnetite (Fe$_3$O$_4$) light excitation couples to the critical fluctuations of the charge order and coherently generates structural modes of the ordered phase above the critical temperature of the Verwey transition. Our findings are obtained by detecting coherent oscillations of the optical constants through ultrafast broadband spectroscopy and analyzing their dependence on temperature. To unveil the coupling between the structural modes and the electronic excitations, at the origin of the Verwey transition, we combine our results from pump-probe experiments with spontaneous Raman scattering data and theoretical calculations of both the phonon dispersion curves and the optical constants. Our methodology represents an effective tool to study the real-time dynamics of critical fluctuations across phase transitions.

~\\
$^{*}$Corresponding author: fabrizio.carbone@epfl.ch.
\end{abstract}

\maketitle

\section*{Introduction}

Magnetite (Fe$_3$O$_4$) is the earliest discovered magnetic material. Nowadays, it is widely used in recording devices and as a catalyst \cite{data,cata}, and its fast reaction to light excitation holds promise for future applications in oxide electronics \cite{de2013speed,randi2016phase}. At room temperature, the ground state of magnetite is half-metallic ferrimagnetic and the crystalline structure is an inverse cubic spinel, in which tetrahedrally coordinated A-sites are occupied only by Fe$^{3+}$ ions, while octahedrally coordinated B-sites are occupied by an equal number of nominal valence Fe$^{3+}$ and Fe$^{2+}$ ions. At the Verwey temperature, T$_\mathrm{V}$ (= 116 K in our sample), important discontinuous structural and electronic modifications take place. In particular, the crystal symmetry lowers from cubic $Fd\bar{3}m$ to monoclinic $Cc$ and the dc conductivity decreases by two orders of magnitude. The insulating state is the result of the long-range charge and orbital order (CO-OO) that emerges among the B-sites in the low-temperature (LT) phase. 

Originally, Verwey and Anderson postulated a purely electronic mechanism for the transformation \cite{verwey1939electronic,anderson1956ordering}, which implies an alternate occupation of the $a$/4-spaced (001)$_\mathrm{c}$ lattice planes (where $a$ is the lattice parameter) by Fe$^{2+}$ and Fe$^{3+}$ ions below T$_\mathrm{V}$, and a random distribution of the latter cations above T$_\mathrm{V}$, allowing for electronic transport. Although the above model proves correct to a first approximation, recently a far more tangled charge and orbital arrangement was refined below T$_\mathrm{V}$, comprised of excess electrons delocalized over chains of three B-type Fe ions, termed trimerons \cite{wright2001long,yamauchi2009ferroelectricity,senn2012charge}. 
The trimerons are examples of molecular polarons in which the electronic distribution is interrelated with the shortening of the Fe-Fe distances. 
Interestingly, extensive research on precursors of the phase transition (PT) substantiates the persistence of local dynamical CO-OO in the form of fluctuating trimeron complexes from T$_\mathrm{V}$ to room temperature \cite{fujii1975study,shapiro1976neutron,yamada1980neutron,schwenk2000charge,seikh2005,reznicek2012magneto,hoesch2013anharmonicity,bosak2014short}. However, the nature of the short-range order in the pre-transition region and the relation between the critical correlations which appear gradually above T$_\mathrm{V}$ and the discontinuous modifications at T$_\mathrm{V}$ remain contentious. Sample dependent effects on the critical phenomena prevent more general considerations \cite{lorenzo2008charge,garcia2009reexamination}.

Optical probes, such as reflectivity \cite{park1998charge}, ellipsometry \cite{randi2016phase}, infrared spectroscopy and spontaneous Raman scattering \cite{gasparov2000infrared,yazdi2013raman} have disclosed important information on the electronic and structural properties of magnetite relevant to the mechanism of the Verwey transition. Transfers of spectral weight among different energy regions of the optical spectrum are signatures of the strongly correlated nature of magnetite \cite{maldague2013coupling}, which manifest themselves both in the discontinuous change of the electronic structure at T$_\mathrm{V}$ and in the precursor variations which progressively preempt the Verwey transition. The absorption structures below 2.5 eV were generally attributed to interband transitions between Fe $3d$ states, although alternative interpretations were proposed, at variance on whether only B- or also A-sites are involved \cite{zhang1991electron,park1998charge,leonov2006electronic,randi2016phase}. The appearance of a rich spectrum of additional infrared- and Raman-active modes below T$_\mathrm{V}$ reflects the symmetry lowering of the crystal structure which characterizes the Verwey transition \cite{gasparov2000infrared,yazdi2013raman}.

The structural transformation of magnetite is a complex process, in which small atomic displacements from the prototype to the distorted structure are the result of the concomitant condensation of multiple phonons of comparable importance, rather than dominant contributions from few critical modes. The details of the distortions accompanying the Verwey transition have been elucidated by means of X-ray and neutron scattering \cite{samuelsen1974low,iizumi1982structure,wright2001long,blasco2011structural,senn2012charge}. In particular, below T$_\mathrm{V}$, subsets of modes at the $\Gamma$-, $\Delta$-, X- and W-point are frozen into the static deformations of the low-symmetry structure. Verwey and Anderson's scenario, and subsequent Cullen and Callen's theory \cite{cullen1970} only account for electron-electron interactions and thus cannot explain the structural transformation. The involvement of the structural degree of freedom in the transition process was first put forward by Yamada, who proposed that the condensation of fluctuations of the charge density coupled to a $\Delta_5$ symmetry phonon is the main driving force for the metal-insulator transition (MIT) \cite{yamada1975charge}. More recent first principle computations suggest that both the $\Delta_5$ and X$_3$ modes are the primary order parameters of the structural transformation, and the strong coupling of the X$_3$ mode with the conduction band electrons primarily causes the insulating state below T$_\mathrm{V}$ \cite{piekarz2006mechanism,piekarz2007origin}.

In Yamada's model, the modes of the charge density are treated as classical Ising variables and the transformation as an order-disorder process. Therefore, the spectrum of the electronic excitations of the order parameter is essentially a relaxation response centered at zero frequency, whose width is determined by the inverse relaxation time. A radically different model considers a band of quasi-one-dimensional fermionic quasiparticles, generated by the correlations in the system, and interprets the PT in terms of a Peierls instability \cite{seo2002}. In this case, one expects a richer spectrum of electronic excitations, namely, quasiparticle transitions across the Peierls gap and a collective mode arising from amplitude fluctuations of the order parameter at finite frequency \cite{Travaglini1983}.

To map the dynamical evolution of the structural modes across the Verwey transition and their interplay with the electronic excitations, we performed out-of-equilibrium optical spectroscopy experiments. We found that the critical fluctuations of the charge-density-wave (CDW) couple to specific ionic motions, giving rise to precursor effects above the critical temperature. In particular, in this paper we show that an optical pump pulse can transiently promote time-dependent electron-phonon interactions in a sample of natural magnetite and generate coherent phonons which, by symmetry, should be forbidden above the charge ordering temperature. The launch of these forbidden phonons needs the assistance of fluctuations of the ordered phase. This is different from previously proposed mechanisms of coherent phonon generation \cite{merlin1997generating,stevens2002coherent} in which a time-dependent force acts on the ions without the assistance of other low-lying modes. 

\textit{Ab initio} calculations of the phonon dispersion curves in the monoclinic structure allow us to identify the collective modes that are impulsively photoexcited with the X$_3$ (13.9-15.6 meV), $\Delta_2$ (18.5-19.6 meV) and T$_\mathrm{2g}$ mode (25.9 meV). Our assignment of the coherent phonons is further complemented by theoretical predictions of the Raman matrix elements (RME). The correspondence between data and computations serves as an anchor point to rediscuss the attribution of the optical absorption peaks at low energy. As a by-product, we clarify the origin of the transitions at optical frequencies debated in literature.
Furthermore, our results provide a measure of the amplitude and persistence time of the critical fluctuations above T$_\mathrm{V}$. 

\section*{Experiments}

A natural single crystal of magnetite was cut and polished along the (110)$_\mathrm{c}$ plane. In the Supplemental Material (SM), we report a thorough characterization by means of resistivity and ac magnetic susceptibility experiments, revealing the typical fingerprints of high-quality single crystals.

Figure 1 displays the steady-state optical conductivity of magnetite in the low energy region, where temperature dependent changes of the spectral weight take place. As T$_\mathrm{V}$ is crossed, an optical gap opens below 0.2 eV, signaling the onset of the insulating phase. 
Three main features are respectively centered at 0.6, 2 and 5 eV. 
Based on density functional theory (DFT) calculations \cite{zhang1991electron}, Park and co-workers first attributed such features to intersite transitions between B-type Fe t$_\mathrm{2g}$ states, between A- and B-sites and the charge-transfer (CT) between O $2p$ and Fe $3d$ states, respectively \cite{park1998charge}. 
Advanced DFT+U calculations revised the latter assignment, indicating that the absorption feature at 2 eV does not involve A-type Fe sites, but only B-type Fe sites with different valence states \cite{leonov2006electronic}. Nevertheless, no meaningful intensity was predicted for this peak, which makes the assignment put forth in Ref.~\citenum{leonov2006electronic} open for discussion.
 
In our pump-probe experiments, the photoexcitation energy was tuned to the CT region (3.1 eV) and the probe band spanned the range of the $d$-$d$ transitions (1.7-2.9 eV). The sample was fixed to the copper cold-finger of a closed-cycle He cryostat, with a base pressure of 10$^{-8}$ mbar. The output of an amplified Ti:Sapphire laser delivering 50 fs pulses centered at 1.55 eV at a repetition rate of 6 kHz was split into a pump and a probe beam. The pump beam was used to photoexcite the material, after doubling the fundamental frequency in a beta-barium borate crystal, whereas the probe beam was focused on a CaF$_2$ crystal to generate broadband pulses from 1.7 to 2.9 eV. The reflectivity spectra were recorded by a fiber-coupled spectrometer. A chopper modulated the train of pump pulses at 3 kHz, so that at every delay time the difference between the pumped and unpumped reflectivity spectrum was measured, yielding a sensitivity down to 10$^{-4}$. Further details of the set-up are described in Refs. \citenum{mann2015probing} and \citenum{mansart2013coupling}. 
The delay time between the two pulses was varied to obtain a sequence of time points representative of the photoinduced dynamics of the material. Spontaneous Raman scattering at 2.4 and 3.1 eV excitation energies was also carried out to 
discriminate between equilibrium and non-equilibrium effects (see SM for further details).

\section*{Results}

\subsection*{Ultrafast broadband spectroscopy}

The spectrum of the transient reflectivity as a function of  pump-probe delay time, $\Delta$R/R~(E,~t), at 10 K and $\sim$1 mJ/cm$^2$ fluence is shown in Fig. 2a.
The $\Delta$R/R temporal traces averaged over 0.2 eV wide energy intervals are displayed in Fig. 2b.
Damped coherent oscillations lasting over 1.5 ps superimposed to a multi-exponential decay 
clearly emerge 
in all the time traces of Fig. 2b and in the colormap of Fig. 2a.
To analyze these oscillations, the temporal trace averaged between 2.0 and 2.3 eV probe photon energy is fitted with the sum of multi-exponential decays and two distinct damped coherent oscillations, convolved with a Gaussian response function. The data and the fitting functions are displayed in Fig. 2c.
The beating between the two oscillations shown in Fig. 2d is thereby isolated.

The same fit analysis is repeated at different temperatures and the same fluence.
The temperature dependence of all oscillations is displayed in Fig. 3a--d.
Figure 3e--h shows the evolution of the oscillation amplitude, as obtained by Fourier transforming the time traces.
According to the fit results, the energies of the two main modes are respectively in the 13.9-15.6 meV and 18.5-19.6 meV ranges. 

As illustrated in Fig. 3b and 3d, in the cubic phase, $\Delta$R/R also shows a weak coherent oscillation corresponding to an energy of 25.9 meV, visible in a different spectral range ($<$ 1.9 eV). 
Remarkably, although the intensity of the 13.9-15.6 meV and 18.5-19.6 meV oscillations decreases abruptly above T$_\mathrm{V}$, it does not vanish up to 140 K (see Fig. 3e and 3g). 
Instead, the 25.9 meV oscillation is only visible in the high-temperature (HT) phase and progressively weakens with increasing temperature (see Fig. 3f and 3h).

Next, in Fig. 4a--d, we show the dependence of $\Delta$R/R on the fluence at 120 K, right above the transition temperature. 
As evidenced in Fig. 4d, the 13.9-15.6 meV and 18.5-19.6 meV oscillations are completely quenched with increasing fluence above $\sim$1 mJ/cm$^2$. 
In contrast, the intensity of the 25.9 meV oscillation increases with fluence and saturates above $\sim$4 mJ/cm$^2$. 

\subsection*{Mode assignment and phonon calculations in $P2/c$ symmetry}

To assign these modes, a direct comparison of the FT of the coherent oscillations and the spontaneous Raman spectra with the theoretical predictions is carried out in Fig. 5a--d. 
The phonon dispersion curves obtained from \textit{ab initio} calculations in the cubic phase below 27 meV are represented by red lines in Fig. 5c.
Figure 5d shows the spontaneous Raman spectra at room temperature for 2.4 and 3.1 eV excitation energies, and the FT of the coherent oscillations detected in our pump-probe experiments. 
The only Raman-active phonon at the zone center in the relevant energy region is the T$_\mathrm{2g}$ mode. 

In the monoclinic phase, the structural PT causes a quadrupling of the unit cell, resulting in new phonon branches, as shown by the blue lines in Fig. 5b. 
As a consequence, various modes are folded to the zone center and become potentially observable by visible-light Raman scattering, as illustrated by the horizontal lines in Fig. 5a. 
Figure 5a also displays the spontaneous Raman spectra of our single crystal at 5 K for 2.4 and 3.1 eV excitation energies, and their time-resolved counterpart at 10 K. The energy ranges of the coherent oscillations as inferred from the fit analysis are highlighted by shaded areas.

The phonon dispersion curves were calculated in the monoclinic structure (with the approximate symmetry $P2/c$) using the {\it ab initio} direct method \cite{parlinski1997direct,phonon2013}. The electronic structure and atomic positions were optimized using the projector augmented-wave \cite{blochl1994PAW} and generalized gradient approximation (GGA) \cite{pardew1996GGA} implemented in the VASP program \cite{kresse1996VASP}.
The strong electron interactions in the Fe($3d$) states were included within the GGA+U method \cite{ldau}. 
The Hellmann-Feynman forces were calculated by displacing all non-equivalent atoms from their equilibrium positions and the	force-constant matrix elements were obtained in the 112-atom supercell. The phonon dispersions along the high-symmetry directions in the reciprocal space were calculated by the diagonalization of the dynamical matrix. The phonon dispersions in the HT cubic phase were studied using the same approach and discussed in Refs. \citenum{piekarz2006mechanism} and \citenum{piekarz2007origin}.

We observed that for crossed pump and probe polarizations the signs of the oscillations visible in the monoclinic phase do not change, in contrast to the behavior typical of B$_\mathrm{g}$ symmetry phonons. Therefore, we rule out the LT B$_\mathrm{g}$ modes from the potential candidates for the assignment of these oscillations. On the basis of the comparison presented in Fig.~5, the 13.9-15.6 meV and 18.5-19.6 meV oscillations are ascribed to the LT A$_\mathrm{g}$ modes that are closest in energy, which originate from respectively the lowest-energy X$_3$ and $\Delta_2$ modes in the HT phase. Instead, the 25.9~meV oscillation observed above T$_\mathrm{V}$ is unambiguously identified by its energy with the T$_\mathrm{2g}$ mode of the cubic structure. 
	
Let us note that the peaks in the FT of the coherent oscillations extend in a broad energy range from 11.5 to 21.0 meV. The occurrence of oscillation dephasing and energy renomalization from anharmonic effects may explain the large width of such features and the lower energy of the oscillations compared to their spontaneous Raman counterparts. Our calculations in the approximate symmetry $P2/c$ do not fully account for the effects of CO-OO on the phonon energies, which are thus underestimated, incidentally resulting in a better agreement with the time-resolved data than with the spontaneous Raman spectra. 

The cubic $\Delta_5$ mode gives rise to a monoclinic A$_\mathrm{g}$ mode in the energy range intermediate between the 13.9-15.6 meV and 18.5-19.6 meV oscillations, which may cast doubts on the above assignment based on energy considerations. Such doubts will be dispelled in the following. 

\subsection*{RME analysis and optical constant calculations in $P2/c$ symmetry}

To corroborate our assignments of the oscillations, we benefit from the broadband nature of our optical probe and compare the energy dependence of the experimental RMEs of the oscillations to those obtained from our {\it ab initio} calculations. Singular value decomposition (SVD) of the $\Delta$R/R matrix was performed to separate the energy dependence of the incoherent relaxation of the excitations from that of the collective coherent modes corresponding to the oscillations. Qualitative information on the coupling between structural distortions and electronic excitations was thereby gained. Our algorithm decomposed the $\Delta$R/R matrix into the sum of outer products between orthonormal spectral and temporal vectors, also termed canonical traces, weighted by singular values. The sum was truncated to account only for the physically relevant terms, thus reducing the noise level. The canonical traces were fitted with model functions (see Supplementary Fig. 4). By that, as shown in Fig. 6a and 6b, the incoherent and the coherent contribution to $\Delta$R/R were reconstructed. The same analysis combined with a fit using a Lorentz model delivered the energy profiles of the differential permittivity associated with the two oscillations at the maximum displacement from the equilibrium positions, depicted in Fig. 6c--e and in Fig. 7b, in a wider energy range. A thorough description and application of our SVD-based method is available in Ref. \citenum{mann2015probing}.

To obtain the theoretical RMEs, we followed the procedure described in Ref.~\citenum{mann2015probing}. The frequency-dependent dielectric function was calculated for a series of monoclinic unit cells (with approximate symmetry $P2/c$) whose atoms were displaced according to the eigenvectors of the phonon modes in the same 11.5-21.0 meV energy range as the oscillations observed in our experiments. These are the A$_\mathrm{g}$ modes of the monoclinic structure that originate from the cubic X$_3$, $\Delta_5$ and $\Delta_2$ counterparts. We computed the RMEs as the finite differences of the dielectric function associated with the structural distortions in the linear displacement regime. The self-consistent charge density of the electronic ground state was obtained from DFT calculations in the GGA+U approximation implemented in the {\sc quantum espresso} package~\cite{QE_pwscf,ldau}. Subsequently, the frequency-dependent dielectric function was calculated with the linear response method within the random phase approximation using the {\sc yambo} package~\cite{Yambo_ref} (see the SM for computational details). For values of the on-site Coulomb repulsion U and the Hund's exchange coupling J of respectively 4 eV and 1 eV, there is qualitative similarity between the computed and experimental RMEs for the X$_3$ and $\Delta_2$ modes, whereas they are in complete disagreement for the $\Delta_5$ mode (see Fig. 6c--e). Notice that the computed RMEs depend on the energy locations of the transitions which, in general, are only approximately predicted by \textit{ab initio} calculations. This may explain the relative energy shift of data and predictions for the X$_3$ mode (see Fig. 6c), but not the inconsistency of the theoretical RME for the $\Delta_5$ mode with both measured RMEs (see Fig. 6d). Thus, our RME analysis provides an anchor point to our assignment of the oscillations in the foregoing following from energy considerations alone.

In addition, as illustrated in Fig. 7a, our calculations qualitatively reproduce the characteristic features of the optical conductivity at 15 K, except for the position of the lowest-energy peak. We note that the agreement with the experimental data did not improve for other possible choices of the U parameter (see Supplementary Fig. 5). Considering the reasonable consistency of our predictions to the experimental RMEs in the probed energy range, we re-examine the absorption structure of magnetite in the LT phase to better substantiate the assignment of the interband transitions. As illustrated in Fig. 7a, the total spectrum was decomposed into the main contributions from different interband transitions (our complete analysis is presented in Supplementary Fig. 6). Individual electronic transitions were discriminated based on the symmetry of the energy bands corresponding to different ions, summarized in Fig. 7c. Three separate components dominate the debated feature around 2 eV (represented with different colors in Fig. 7a), namely, the CT between (i) the minority-spin t$_\mathrm{2g}$ and e$_\mathrm{g}$ states of respectively nominal valence Fe$^{2+}$ and Fe$^{3+}$ B-sites, (ii) the minority-spin t$_\mathrm{2g}$ states of the same ions and (iii) the top of the highest filled band and the bottom of the lowest empty band 
formed by respectively B- and A-type Fe ions in the majority-spin channel. Earlier calculations~\cite{leonov2006electronic} attributed the peak centered around 2 eV only to the first of the above three excitations purely based on energy considerations, albeit with negligible calculated intensity. In contrast, our assignment relies on the computed spectral weights, which reveal the additional involvement of the A-sites. Furthermore, we found that the excitation of the t$_\mathrm{2g}$ electrons across the bandgap extends in the energy region relevant to the 2 eV feature in the form of a satellite peak.

\subsection*{CDW-assisted excitation of the LT modes above T$_\mathrm{V}$}

In the HT phase, the X$_3$ and $\Delta_2$ modes are not at the zone center, thus they should not be accessible by low-momentum-transfer experiments such as Raman scattering (see Fig. 5c). Here, we show that the forbidden modes can become active with the assistance of CDW fluctuations. Such precursor effects can be expected in PTs, except for transitions which are mean-field-like  or 
have a strong first-order character. In magnetite, such fluctuations have been demonstrated to persist above T$_\mathrm{V}$ \cite{bosak2014short,fujii1975study} and provide the (0,0,1) and (0,0,$\frac{1}{2}$) wave-vector components (in units of $2\pi/a$) for momentum conservation, in the Raman process pictorially represented in Fig. 8.

We model the light pulses as a time-dependent electric field ${\bf E}$, with an oscillation in the range of visible light frequencies, corresponding to the laser central frequency $\omega_{L}$, and a slowly-varying modulation $ \bm{\mathcal{E}}$, representing the pulse shape, ${\bf E}(t) =  \bm{\mathcal{E}}(t) e^{i\omega_{L} t}$. We introduce boson creation ($a^\dagger_{{\bf q}\lambda}$) and destruction ($a_{{\bf q}\lambda}$) operators for phonons of branch $\lambda$ and momentum ${\bf q}$, and the corresponding canonical coordinate, $Q_{{\bf q} \lambda} =\left({\hbar}/{2\omega_{{\bf q}\lambda}} \right)^{1/2}(a_{{\bf q} \lambda}+a^\dagger_{-{\bf q} \lambda})$. The order parameter can be either a bosonic or a classical variable, depending on whether one considers the PT as a Peierls instability, with quantum fluctuation effects \cite{lee1974}, or an order-disorder process, described by a $\phi^4$ theory, with a central peak-like dynamical response \cite{chaikin2000}.

The time-dependent perturbation introduced by the pump electric field can be expressed by the following two-mode Raman Hamiltonian, which couples CDW fluctuations and a phonon mode, thus acting as a time-dependent electron-phonon interaction controlled by the photoexcitation,
\begin{equation}
\label{eq:hram}
H_R=\sum_{{\bf q} \lambda} g_{{\bf q} \lambda}(t) \rho_{{\bf q}} Q_{-{\bf q} \lambda}
\end{equation}
with 
\begin{equation}
\label{eq:k12}
 g_{{\bf q} \lambda}(t)=-\frac{V}{4} \bm{\mathcal{E}}(t) \frac{\partial^2 \bm{\chi}}{\partial  \rho_{{\bf q}}\partial  Q_{-{\bf q} \lambda} }(\omega_L) \bm{\mathcal{E}}^*(t).
\end{equation}
Here, $\rho_{{\bf q}}$ is the canonical coordinate for the charge fluctuations with momentum ${\bf q}$, $V$ is the sample volume, and $  {\partial^2 \bm{\chi}}/{\partial  \rho_{{\bf q}}\partial  Q_{-{\bf q} \lambda} }$ is the Raman tensor, expressed as the second derivative of the dielectric susceptibility evaluated at the pump frequency, $\omega_L$. Such Raman tensor is a straightforward generalization of those in Refs. \citenum{mansart2013coupling,cardona1982light,merlin1997generating}. For simplicity, we assume that bare electron-phonon interactions are already incorporated in the definitions of the fluctuation operators, which are \ quasiparticle operators. For the charge fluctuations, the derivative can be evaluated with the use of a Lagrange multiplier, as in Ref.~\cite{mansart2013coupling}.

The response produced by the above interaction can be computed as follows, using the Kubo formula, in the response regime linear in fluence,
\begin{eqnarray}
  \label{eq:kubo3dif}
&&\langle \rho_{{\bf q}}Q_{-{\bf q} \lambda}\rangle(t)=-\frac{i}{\hbar}\int_{-\infty}^t dt'\langle[\rho_{{\bf q}}(t)Q_{-{\bf
  q} \lambda}(t) , Q_{{\bf q} \lambda}(t')\rho_{-{\bf q}}(t') ]\rangle  g_{{\bf q}}(t')\nonumber\\
&&=\int_{-\infty}^{t} dt' R(t-t')  g_{{\bf q}}(t').
\end{eqnarray}
where $\langle ... \rangle$ indicate thermal and quantum averages. If charge fluctuations behave as a classical order parameter, $\rho_{{\bf q}}$ can be taken out of the commutator, yielding, 
 \begin{eqnarray}
   \label{eq:respcla}
&& \langle \rho_{{\bf q}}Q_{-{\bf q} }\rangle(t)=\\
&& -\int_{-\infty}^t dt' 
\langle\rho_{{\bf q}}(t) \rho_{-{\bf q}}(t')\rangle 
\frac{e^{-(t-t')/\tau_{ph}}\sin[\omega_{{\bf q}} (t-t')]} {\omega_{{\bf q}}} 
 g_{{\bf q}}(t')\nonumber
 \end{eqnarray}
where $\tau_{ph}$ is a phenomenological phonon lifetime and $\omega_{{\bf q}}$ is the phonon frequency. For notational simplicity, we drop the phonon branch index $\lambda$ and consider a single phonon. 

An impulsive form of $g_{{\bf q}}(t)$ produces sine-like oscillations of the charge-ion correlation, with an initial strength given by the FT of the squared amplitude of the order parameter fluctuations, $ \langle\rho_{{\bf q}}(0) \rho_{-{\bf q}}(0)\rangle $, and an envelope, $ \langle\rho_{{\bf q}}(t) \rho_{-{\bf q}}(t')\rangle $, which carries information on their time correlation. In the case of a step form of $g_{{\bf q}}(t)$, the oscillation becomes cosine-like. Our description represents a generalization of the mechanism proposed in Ref. \citenum{stevens2002coherent} which can be referred to as fluctuation-assisted stimulated Raman scattering (FASRS). 

The launch of the excitation modulates the optical properties through their dependence on the dielectric susceptibility,
\begin{eqnarray}
  \label{eq:dchi}
\delta \bm{\chi}(\omega,t)&=&\sum_{\bf q} \frac{\partial^2 \bm{\chi}}{\partial  \rho_{{\bf q}}\partial  Q_{-{\bf q}} }(\omega) \langle \rho_{{\bf q}}Q_{-{\bf q} }\rangle(t).
\end{eqnarray}

To apply the FASRS mechanism to the present case, we assume a CDW instability occurring at the X-point of the Brillouin zone (BZ), corresponding to a charge stacking along the $z$ direction, described by $\cos(2\pi z/a)$, which yields a charge occupancy (with respect to the average) $-\delta$, 0, $\delta$, 0 for the B-type Fe planes separated by $a/4$. One can additionally consider CDW fluctuations with twice the periodicity, which can assist the launch of modes at the $\Delta$-point of the BZ, and can be described with the same method. 

The dominant CO in real space, with ${\bf Q}$=(0,0,$2\pi/a$)  wave-vector, corresponding to the X-point of the BZ, can be described as, 
$$
\delta \rho({\bf r})= \phi_{1}({\bf r}) \cos({\bf Q}.{\bf r})+ \phi_{2}({\bf r}) \sin({\bf Q}.{\bf r}).
$$
where  the trigonometric factors account for the rapidly varying part of the CO, whereas $(\phi_{1},\phi_{2})$ is a smoothly varying real order parameter. Expanding the energy in a power series of the order parameter and of its gradients yields a coarse-grained Landau energy functional, describing the PT \cite{chaikin2000}.   

Close to the ordering temperature, critical fluctuations of the density appear strongly peaked at the critical wave-vector and we can approximate the smooth momentum dependent quantities in Eq.~\eqref{eq:dchi} to their value at the ordering wave-vector. 
The response reads, 
 \begin{eqnarray}
   \label{eq:dchi2}
&&\delta \bm{\chi}(\omega,t)=  -\frac{\partial^2 \bm{\chi}}{\partial  \rho_{{\bf Q}}\partial  Q_{-{\bf Q}} }(\omega)\frac{N^2}{2}\times\\
&& \int_{-\infty}^t dt' \sum_{\alpha=1,2} \langle \phi_{\alpha 0}(t) \phi_{\alpha0}(t') \rangle 
  \frac{e^{-(t-t')/\tau_{ph}}\sin[\omega_{{\bf Q}} (t-t')]} {\omega_{{\bf Q}}} 
g_{{\bf Q}}(t')   \nonumber
 \end{eqnarray}
with $N$ the number of sites and $  \langle \phi_{\alpha 0}(t) \phi_{\alpha0}(t') \rangle $ the order parameter autocorrelation at ${\bf r}=0$. Neglecting the fluctuations in Landau theory, the autocorrelation is proportional to the square of the order parameter, which increases linearly in T$_\mathrm{V}$-T with decreasing temperature, as schematically illustrated in Fig. 9 (blue line). This corresponds to the notion that the modes become active in the broken symmetry phase upon folding from the boundary to the center of the BZ. If the fluctuations are taken into account, the results are different. The equal-time correlation determines the initial amplitude of the oscillations. If one computes the correlation at the Gaussian level, one obtains the red line in Fig. 9, showing that the fluctuations make the modes active even above T$_\mathrm{V}$. Furthermore, if the oscillations are shorter than the phonon lifetime, it means that the decay is dominated by 
the factor $ \langle \phi_{\alpha 0}(t) \phi_{\alpha0}(t') \rangle $ in Eq.~\eqref{eq:dchi2} and one obtains information on the dynamics of the order parameter. For a weakly first-order transition, as in the present case, we expect a small jump of the intensity at the critical point, but a qualitatively similar behavior.

\section*{Discussion}

Over the last decades, revived attention has been focused on the precursors of the PT which appear far above the critical temperature. For instance, disk-like diffuse scattering anticipates the Verwey transition starting from room temperature, as a result of dynamical correlations among trimeron complexes, sustained in the cubic phase \cite{shapiro1976neutron,yamada1980neutron,bosak2014short}. Recently, a lattice dynamical study found anomalies of transverse-acoustic phonons at incommensurate wave-vectors, which originate from their strong coupling to the same fluctuations of the charge density \cite{hoesch2013anharmonicity}. In general, the experimental results provide extensive evidence to the thermal population of degenerate excited state of short-range order prior to the Verwey transition. Degeneracy originates from the equivalent orientations of the order pattern in cubic symmetry, similar to crystal microtwinning in the monoclinic phase, but on a nanometer scale and in dynamical conditions. 

Within the same framework, the excitation and detection of two modes of the monoclinic structure above the critical temperature in our pump-probe experiments can be rationalized considering the effects of critical fluctuations and provide information on their dynamics. Eqs.~\eqref{eq:respcla} and ~\eqref{eq:dchi2} express the fact that, in the presence of a long-lived CDW fluctuation, a lattice vibration with finite momentum can be excited by light, with a coherence time limited by the persistence time of the CDW fluctuation. Practically, the frequency of the modes does not change across the critical region. This means that the modes are rendered active by classical fluctuations of the order parameter, as expected in Yamada's model, consistent with the central peak observed in inelastic neutron scattering \cite{fujii1975study,shapiro1976neutron,yamada1980neutron,bosak2014short}. Instead, if a Peierls mechanism was playing a significant role, one would expect that the electronic spectrum of the modes affects the response. For example, if one approximates the electronic spectrum to a single pole at $\omega_\mathrm{CDW}$, the oscillations should appear at $\omega_{{\bf Q}}\pm\omega_\mathrm{CDW}$, with a strong energy dependence on temperature, which we do not observe.

By means of our SVD algorithm, we were able to determine the experimental profiles of the RMEs as a function of probe photon energy. A comparison with \textit{ab initio} calculations allowed us to unambiguously identify the modes. To the best of our knowledge, this assignment methodology is introduced here for the first time and has the advantage to rely on the phonon eigenvector instead of the phonon energy, thus providing more precise and specific information than other optical approaches. 

Our time-resolved experiments directly access the decay time of the coherent oscillations. According to Eq. \ref{eq:dchi2}, this is limited by the phonon lifetime $\tau_{ph}$ or the persistence time of the CDW fluctuations, whatever is shorter. If the fluctuations of the order parameter are the limiting factor, we expect that the lineshapes of the modes in the energy domain become broader above the critical temperature, as it is indeed the case (see Fig. 3e). Preliminary results from our inelastic neutron scattering \cite{borroni2016experimental} and spontaneous Raman scattering experiments (see Supplementary Fig. 3) indicate that the phonon lifetimes ($>$ 600 fs) are significantly larger than the time constants of the coherent oscillations in accord with this result. Therefore, our measurements provide an estimate of the relaxation time of the fluctuations of the order parameter in the 130-140 K pre-transition region, $\tau_{CDW} \sim 240$ fs.

Gaussian fluctuations result in a cusp anomaly of the oscillation amplitude at the critical temperature (see Fig. 9). Unfortunately, our temperature and energy resolution are not high enough to test this hypothesis. In any case, the persistence of fluctuations above the ordering temperature (see Fig. 3g) is in qualitative agreement with the prediction of the fluctuation model. The intensity drop with increasing fluence for the $\Delta_2$ mode (see Fig. 4d) is probably due to temperature detuning from the critical region, and represents a hint of the fast decrease in intensity above the ordering temperature expected from the fluctuation model.

In contrast to our non-equilibrium experiments, in our spontaneous Raman measurements at 2.4 and 3.1 eV excitation energies, no forbidden mode was detected below 27 meV in the disordered phase (see Supplementary Fig. 3). This is indicative of the different sensitivity of the two techniques to low-energy modes, but could also suggest that non-linear effects enhance the coherent response in our pump-probe experiments. One can easily check that the time-dependent electron-phonon interaction promotes a transient stabilization of the ordered phase, which is equivalent to an increase of the effective ordering temperature in steady-state conditions. Possibly, the above effect amplifies the coherent response, by inducing a breaking of symmetry limited in time and confined in space. Experimentally, to validate this hypothesis, one would need to carry out a thorough study of the fluence dependence of the coherent response in the proximity of the critical temperature. Such an experiment requires accurate temperature control across the specific heat anomaly in the critical region and high enough signal-to-noise ratio to detect small oscillations at low fluences. 

\section*{Conclusions}

Typically, ultrashort laser pulses are used to melt the order parameter of the low-symmetry phase and thus establish the electronic and structural properties of the disordered phase \cite{de2013speed,randi2016phase}. Here, we discussed a more subtle effect, with implications on the coherent control of transition-metal oxides. In particular, we demonstrated that impulsive photoexcitation at 3.1 eV, above the energy onset for O $2p$ -- Fe $3d$ CT transitions, transiently promotes the coupling between lattice vibrations at finite momentum and fluctuations of the electronic order parameter above the ordering temperature of the Verwey transition in magnetite. In particular circumstances, the optical control of the electron-phonon interactions may even induce order. 

Fluctuations of the order parameter are not easy to measure and often require neutron scattering techniques, generally limited in energy resolution. We developed a method able to probe the dependence of the amplitude and persistence time of critical fluctuations on temperature and light excitation. Our method demonstrated the capability to gain insight into the critical softening of fluctuations in correspondence of PTs. Moreover, our approach enabled us to describe the atomic motions associated with the critical fluctuations and their interplay with the electronic degree of freedom. This further elucidated the microscopic mechanism of the MIT and the interpretation of the optical response of magnetite.

\section*{Acknowledgements}

We acknowledge D. Fausti, H. Lee, A. B. Kuzmenko and D. McGrouther for useful discussions, M. Prester and D. Drobac (Institute of Physics, Zagreb) for the AC susceptibility characterization, and A. Pisoni (Laboratory of Physics of Complex Matter, \'{E}cole Polytechnique F\'{e}d\'{e}rale de Lausanne, Lausanne) for the resistivity characterization. Work at LUMES was supported by NCCR MUST and by the ERC starting grant USED258697. J. L. acknowledges financial support by Italian MIUR under project PRIN-RIDEIRON-2012X3YFZ2. V. M. K. and O. V. Y. acknowledge the support from ERC project `TopoMat' (grant No. 306504). First-principles calculations have been performed at the Swiss National Supercomputing Centre (CSCS) under Project No. s675. A. M. O. and P. P. acknowledge the support by Narodowe Centrum Nauki (NCN, National Science Centre, Poland), Project No. 2012/04/A/ST3/00331. D. L. acknowledges the project “IT4Innovations National Supercomputing Center-
LM2015070" and the grant No. 17-27790S of the Grant Agency of the Czech Republic. 

\clearpage
\pagebreak

\clearpage
\begin{center}
\includegraphics[width=0.33\textwidth]{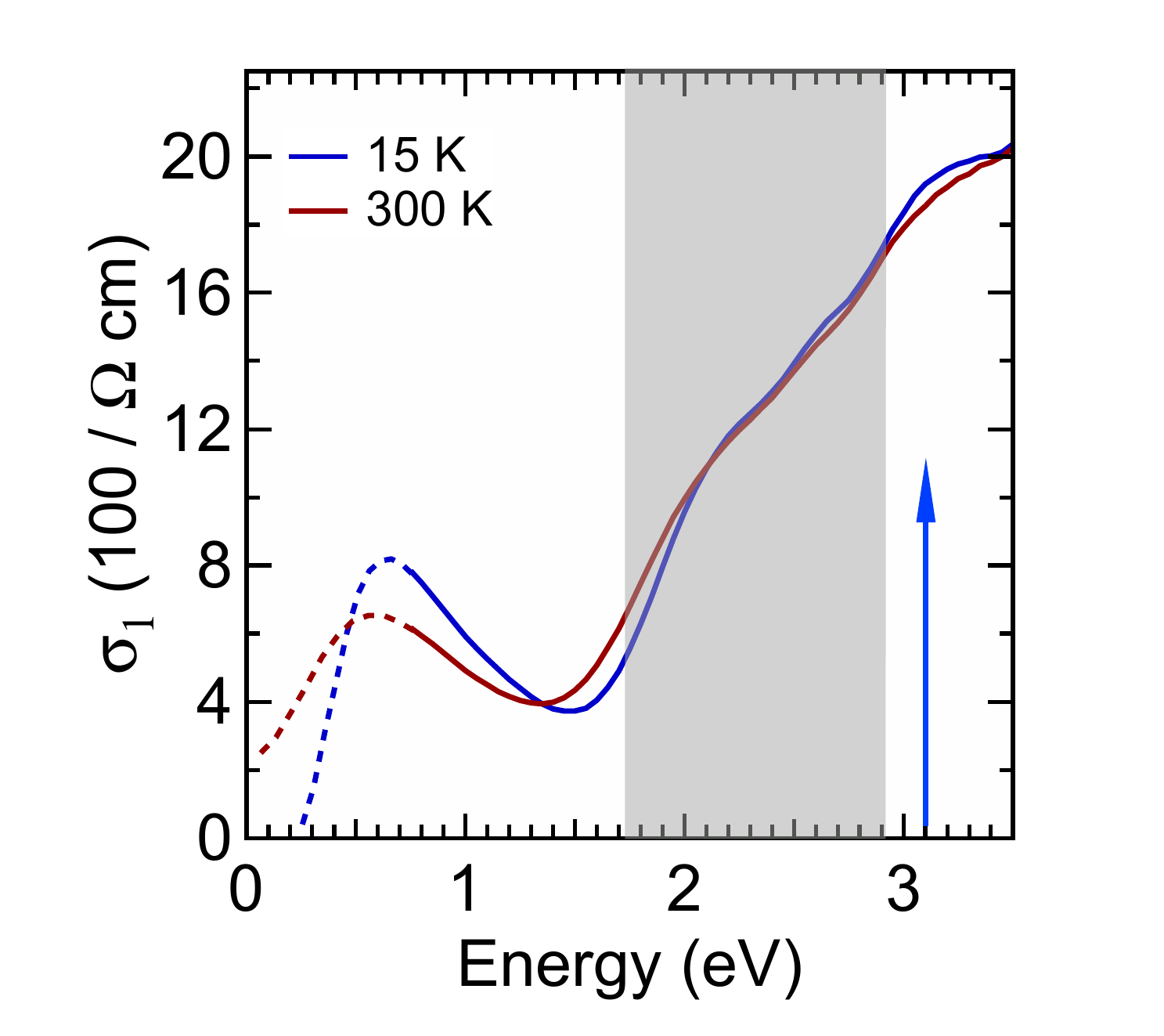}
\end{center}
\setlength{\parindent}{0cm}
\textbf{Figure 1 \textbar \, Steady-state optical properties.} Real part of the optical conductivity at 15 and 300 K from the ellipsometry measurements of Ref. \citenum{randi2016phase}. Lines are dashed where curves are extrapolations based on fits to the data. The blue arrow and the shaded area respectively correspond to the energy of the impulsive photoexcitation and the probed range in our time-resolved experiments.

\clearpage
\begin{center}
\includegraphics[width=\textwidth]{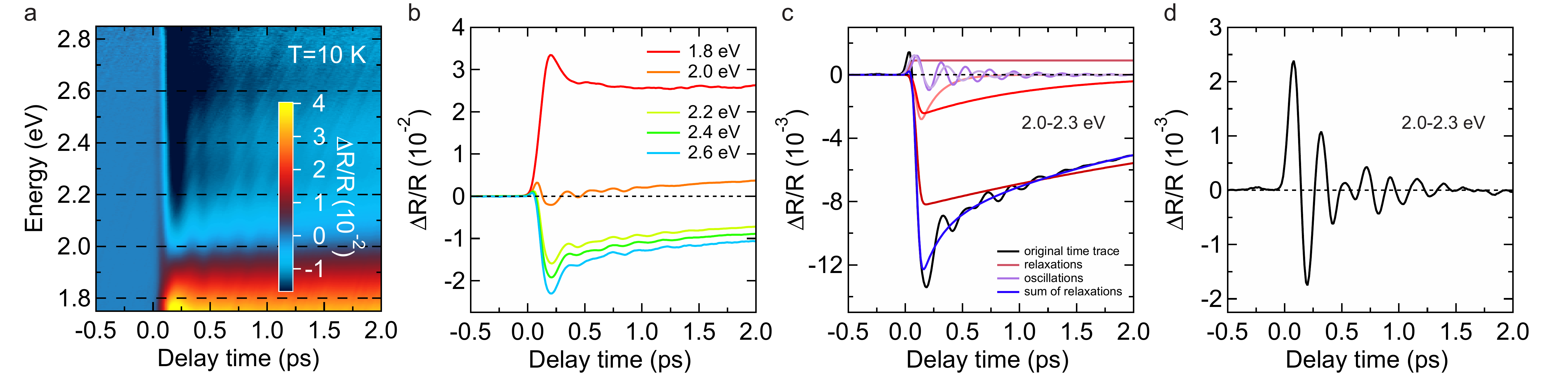}
\end{center}
\textbf{Figure 2 \textbar \, Response to the impulsive photoexcitation in the monoclinic phase.} (\textbf{a}) Color-coded map of differential reflectivity as a function of pump-probe delay time and probe photon energy at 10 K and $\sim$1 mJ/cm$^2$ fluence. (\textbf{b}) $\Delta$R/R time traces averaged over 0.2 eV wide windows centered on the horizontal lines of panel a. (\textbf{c}) Fit analysis of $\Delta$R/R integrated between 2.0 and 2.3 eV. Fitting functions are displayed in different colors. (\textbf{d}) Coherent response isolated through fit analysis of $\Delta$R/R integrated between 2.0 and 2.3 eV.

\clearpage
\begin{center}
\includegraphics[width=0.67\textwidth]{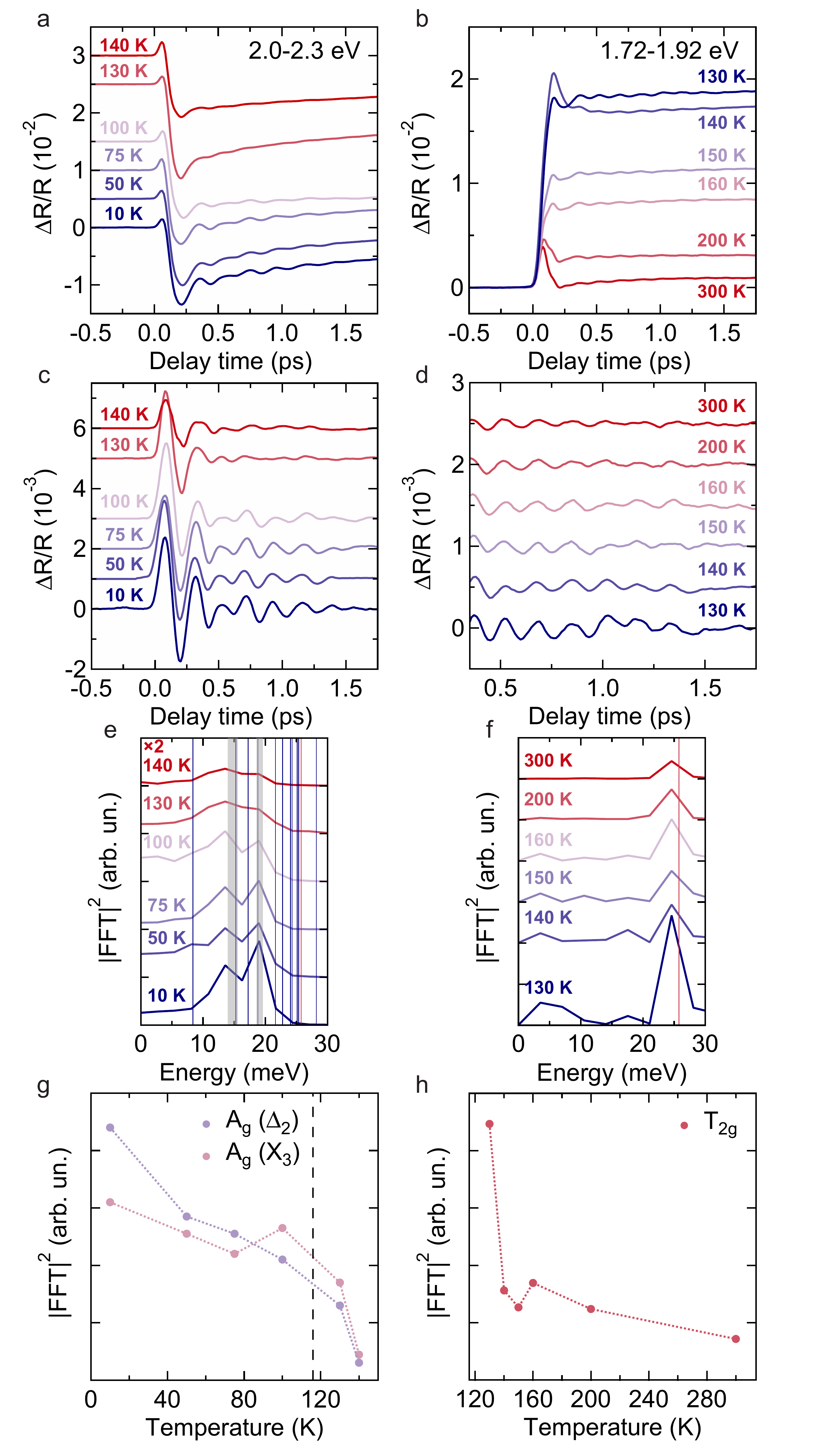}
\end{center}
\clearpage
\setlength{\parindent}{0cm}
\textbf{Figure 3 \textbar \, Temperature dependence of the coherent effects.} (\textbf{a},\textbf{b}) Differential reflectivity as a function of pump-probe delay time averaged over 2.0-2.3 eV and 1.72-1.92 eV energy ranges at different temperatures. (\textbf{c},\textbf{d}) Oscillatory component singled out from the time traces of panel a and b by subtracting the non-oscillatory transient. (\textbf{e},\textbf{f}) Power spectrum of the coherent response from a Fourier transform (FT) of the oscillatory component. Blue (Red) vertical lines indicate the energy of the A$_g$ (T$_\mathrm{2g}$) phonon modes at the $\Gamma$-point in $P2/c$ ($Fd\bar{3}m$) symmetry from \textit{ab initio} calculations. Shaded areas extend over the energy ranges of the beating modes in the LT phase as estimated based on the fit analysis. (\textbf{g},\textbf{h}) Temperature dependence of the peak $\vert$FFT$\vert^2$ corresponding to the oscillations visible by eye. A vertical dashed line denotes T$_\mathrm{V}$.

\clearpage
\begin{center}
\includegraphics[width=0.33\textwidth]{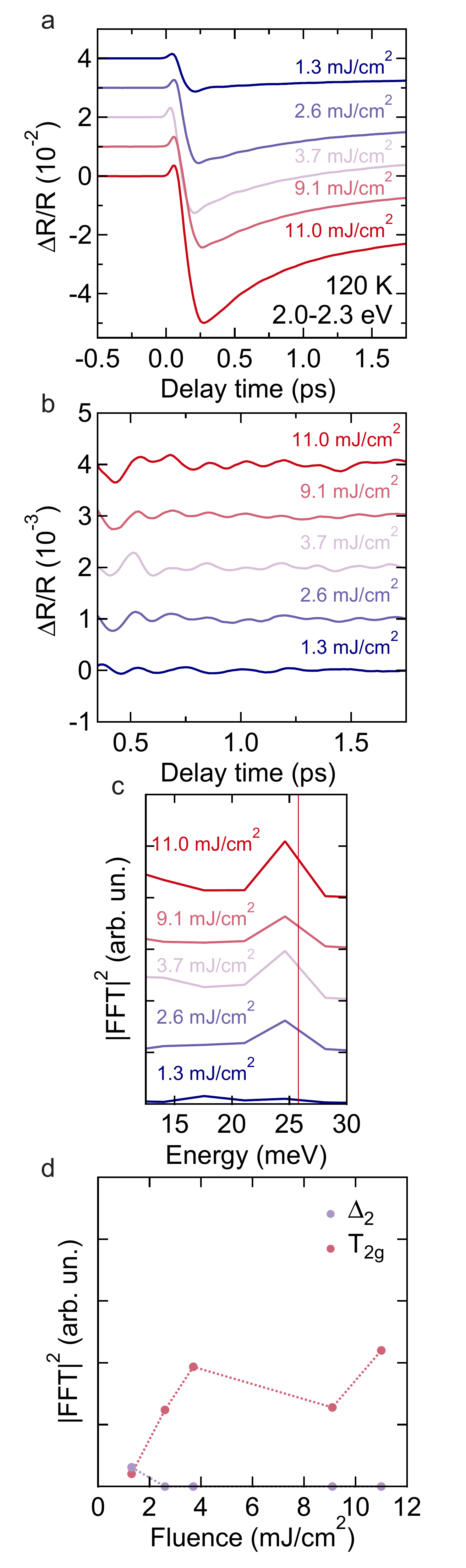}
\end{center}
\clearpage
\setlength{\parindent}{0cm}
\textbf{Figure 4 \textbar \, Fluence dependence of the coherent effects at 120 K.} (\textbf{a}) Differential reflectivity as a function of pump-probe delay time averaged in between 2.0 and 2.3 eV at different fluences. (\textbf{b}) Oscillatory component singled out from the time traces of panel a by subtracting the non-oscillatory transient.  (\textbf{c}) Power spectrum of the coherent response from the FT of the oscillatory component. A red vertical line indicates the energy of the T$_\mathrm{2g}$ phonon mode at the $\Gamma$-point in $Fd\bar{3}m$ symmetry from \textit{ab initio} calculations. (\textbf{d}) Temperature dependence of the peak $\vert$FFT$\vert^2$ corresponding to the oscillations visible by eye.

\clearpage
\begin{center}
\includegraphics[width=\textwidth]{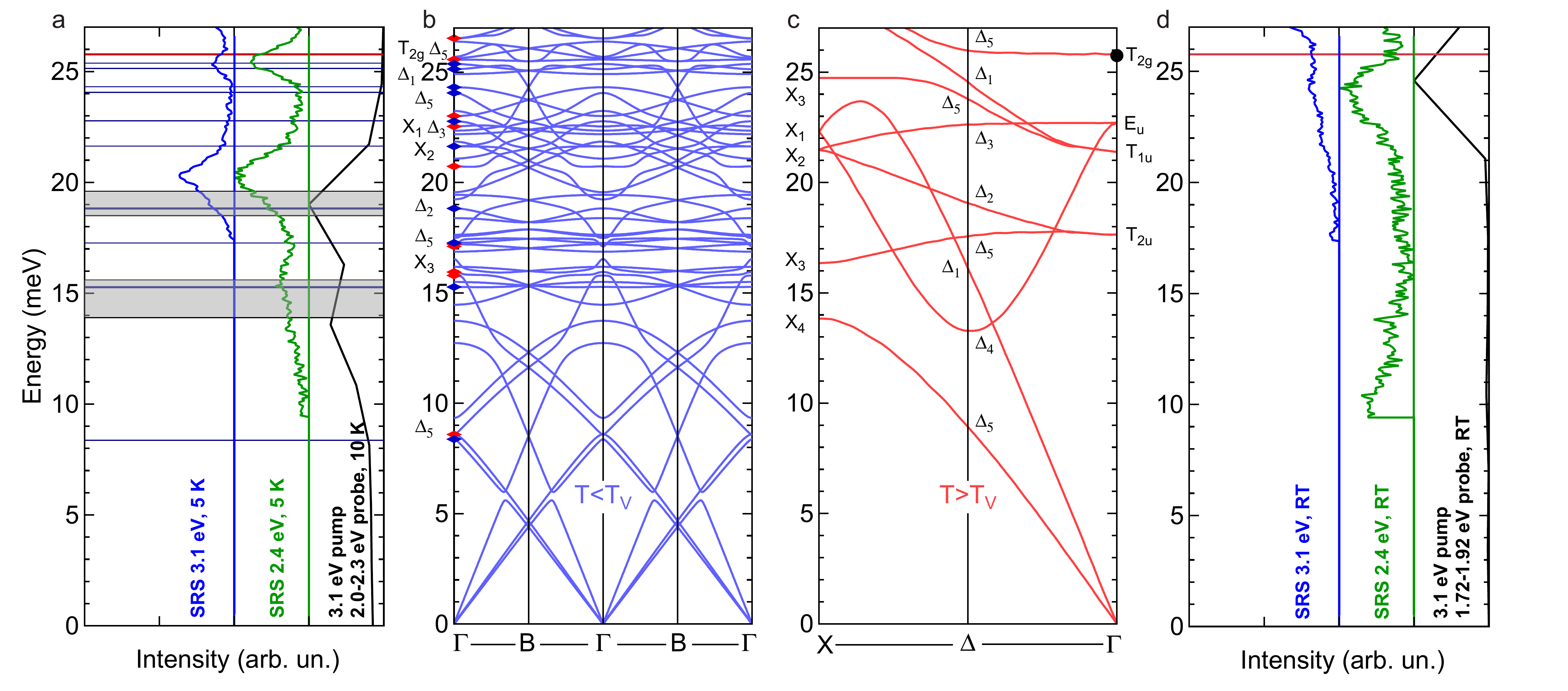}
\end{center}
\textbf{Figure 5 \textbar \, Assignment of the phonon modes in pump-probe and spontaneous Raman experiments.} (\textbf{a},\textbf{d}) Comparison between our spontaneous Raman spectra at different excitation energies (colored lines) and $\vert$FFT$\vert^2$  of the coherent effects in our time-resolved experiments (black line), respectively in the monoclinic and cubic phase. Blue (Red) horizontal lines indicate the energy of the A$_g$ (T$_\mathrm{2g}$) phonon modes at the $\Gamma$-point in $P2/c$ ($Fd\bar{3}m$) symmetry from \textit{ab initio} calculations. Shaded areas extend over the energy ranges of the beating modes in the LT phase as estimated based on the fit analysis. (\textbf{b},\textbf{c}) \textit{Ab initio} calculations of the phonon dispersion curves, respectively in $P2/c$ and $Fd\bar{3}m$ symmetry. Labels indicate the symmetry of the modes at the $\Gamma$-, $\Delta$- and X-point in $Fd\bar{3}m$ symmetry, and the cubic counterparts of the Raman-active modes at the zone center in $P2/c$ symmetry. Blue (Red) diamonds are referred to A$_\mathrm{g}$ (B$_\mathrm{g}$) modes.

\clearpage
\begin{center}
\includegraphics[width=0.975\textwidth]{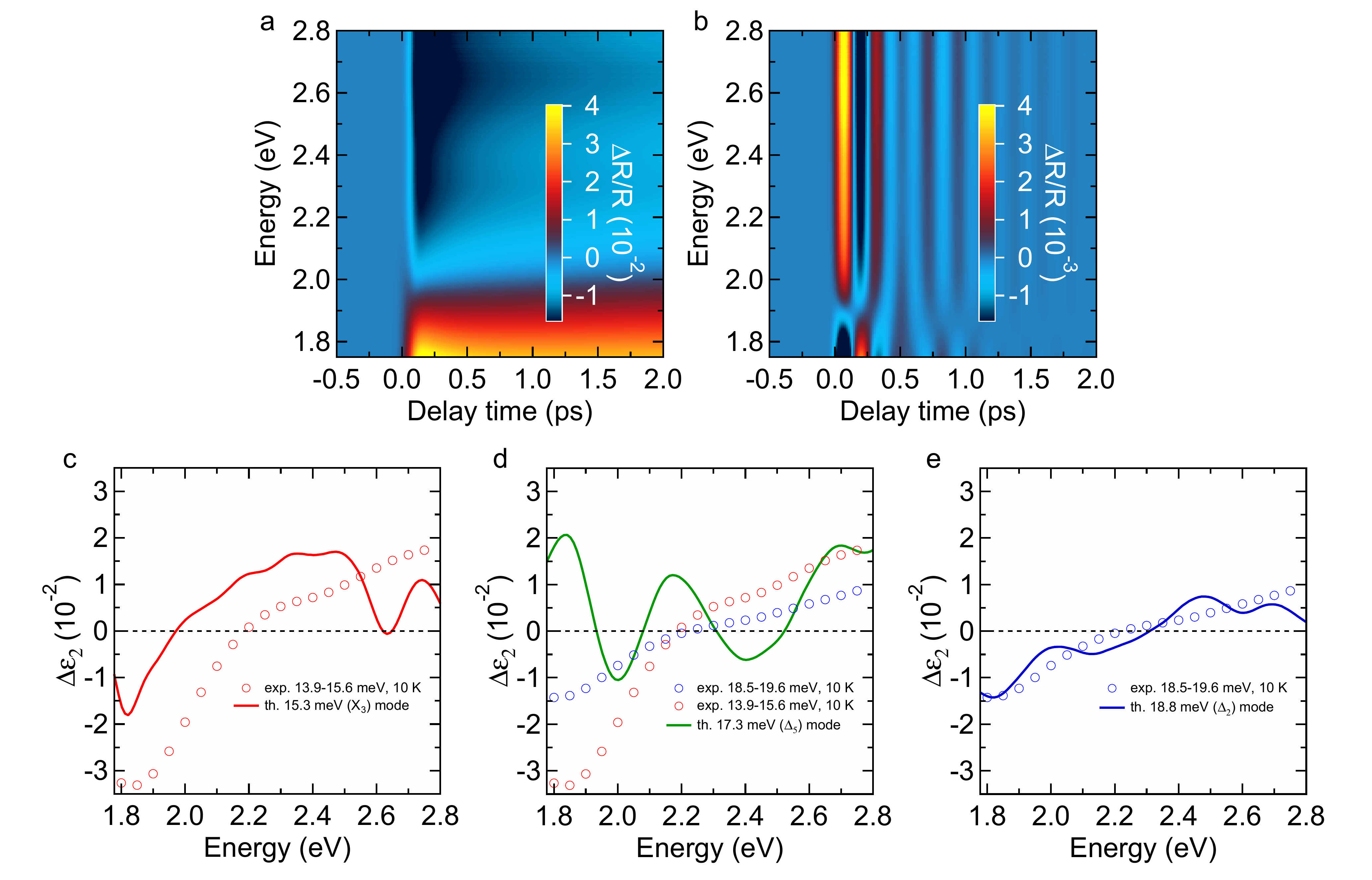}
\end{center}
\textbf{Figure 6 \textbar \, Time and energy dependence of the coherent response.} Reconstruction of the incoherent (\textbf{a}) and the coherent contribution (\textbf{b}) to the $\Delta$R/R matrix at 10 K and $\sim$1 mJ/cm$^2$ fluence through our SVD-based method. (\textbf{c--e}) Energy profiles of the contributions to the imaginary part of the differential permittivity, $\Delta \epsilon_2$, from the oscillations at maximum displacement amplitude at 10 K and $\sim$1 mJ/cm$^2$ (symbols), and calculated RMEs for the A$_\mathrm{g}$ modes in the same energy region, in the monoclinic phase (lines). The cubic counterparts of the modes are specified between parentheses. The computed RMEs correspond to maximum atomic displacements of about $5 \times 10^{-2}$ \AA \: for the 15.3 meV and 18.8 meV modes and $1 \times 10^{-2}$ \AA \: for the 17.3 meV mode. 

\clearpage
\begin{center}
\includegraphics[width=0.65\textwidth]{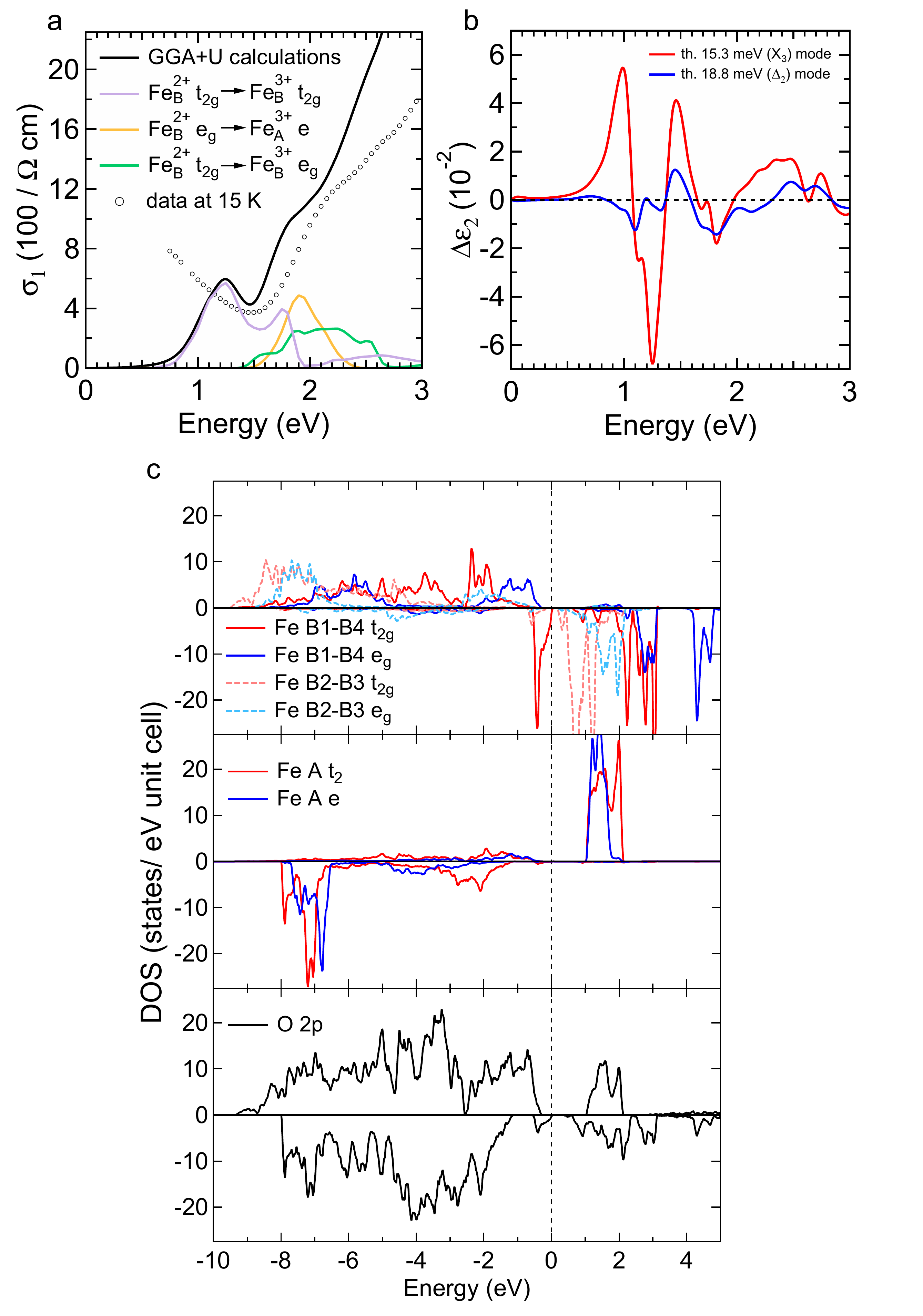}
\end{center}
\setlength{\parindent}{0cm}
\textbf{Figure 7 \textbar \, \textit{Ab initio} calculations of the optical properties.} (\textbf{a}) Comparison between the real part of the optical conductivity at 15 K from the ellipsometry measurements of Ref. \citenum{randi2016phase} and the theoretical predictions from our GGA+U calculations with U = 4 eV and J = 1 eV. The main contributions from different interband transitions to the features around 1.2 and 1.7-2.2 eV are plotted with colored lines and labeled according to the predominant character of the initial and final states. 
(\textbf{b}) Calculated RMEs for the modes corresponding to the oscillations in the low energy region below 3 eV. (\textbf{c}) Calculated partial density of states corresponding to inequivalent Fe and O ions in the $P2/c$ subcell. The notation B1--4 refers to inequivalent positions in the $P2/c$ subcell (see Ref. \citenum{wright2001long}). The Fermi level (dashed line) is set to the top of the valence band.

\clearpage
\begin{center}
\includegraphics[width=0.33\textwidth]{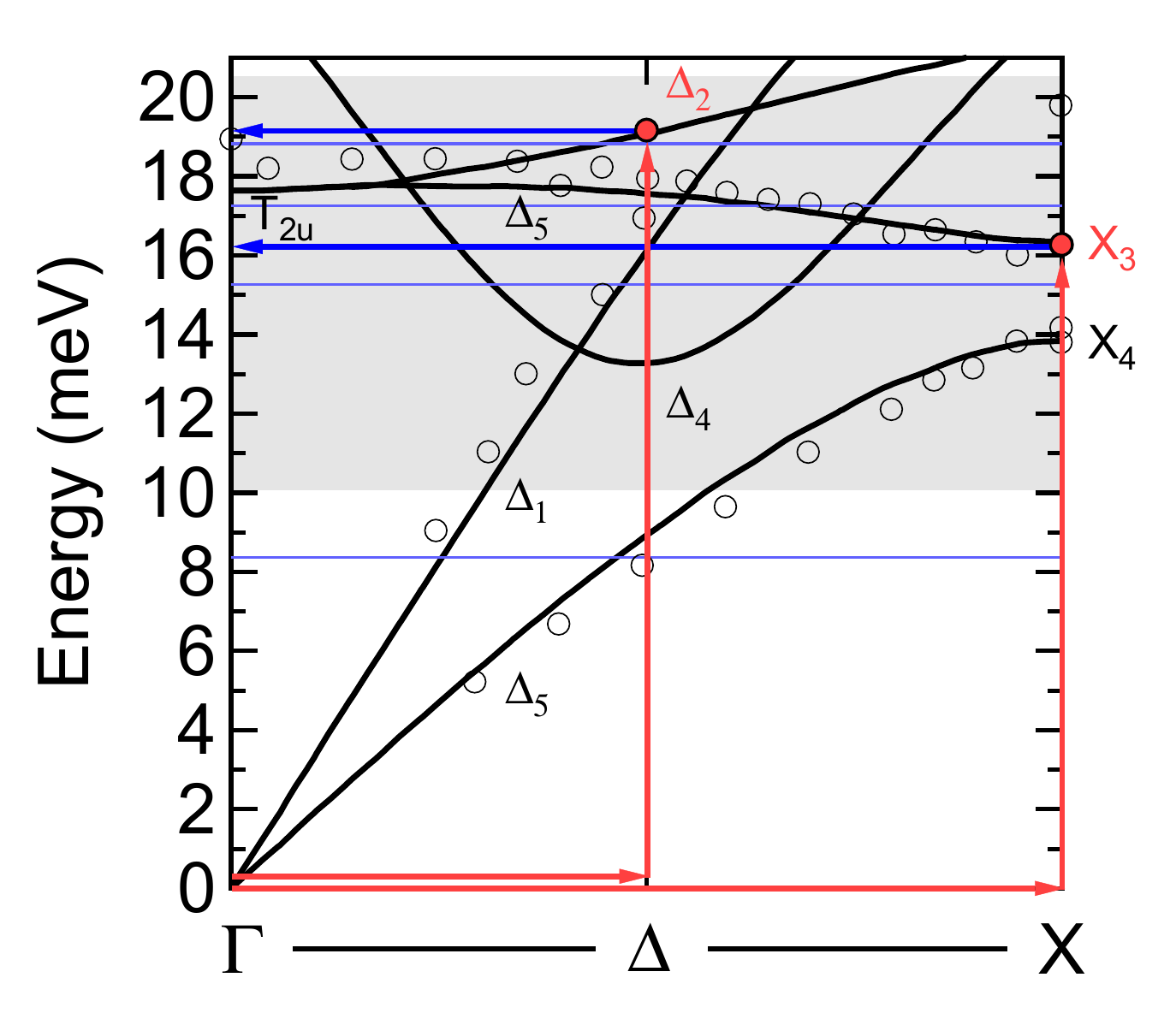}
\end{center}
\textbf{Figure 8 \textbar \, Two-mode photoexcitation mechanism.} Schematic representation of the momentum conservation rule in the photoexcitation mechanism of the X$_3$, $\Delta_2$ and CDW fluctuation modes above T$_\mathrm{V}$. The phonon dispersion curves in the cubic phase as measured by means of inelastic neutron scattering (symbols) \cite{samuelsen1974low} and predicted according to \textit{ab initio} calculations (lines) are shown. The energy and momentum of the phonon and CDW fluctuations are represented by red and blue arrows, respectively. The combination of electronic and vibrational modes allows momentum conservation in the optical process. The shaded area extends over the energy range of the coherent oscillations in the cubic phase,
in the vicinity of T$_\mathrm{V}$, as estimated based on a FT analysis.

\clearpage
\begin{center}
\includegraphics[width=0.36\textwidth]{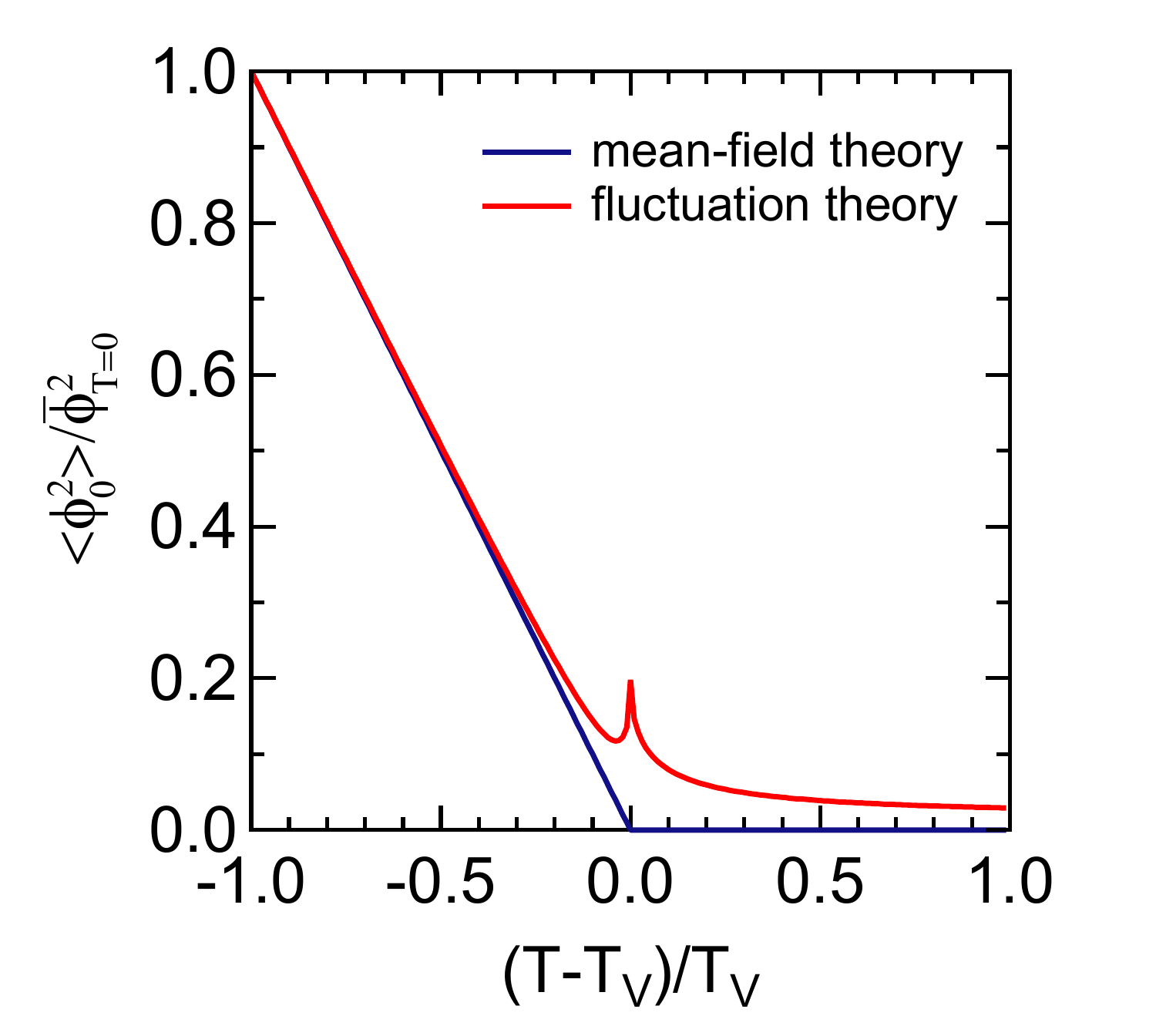}
\end{center}
\textbf{Figure 9 \textbar \, Squared amplitude of the order parameter (fluctuations) as a function of temperature, below (above) the ordering temperature}.
The plotted curves are referred to a one-component $\phi^4$ model at the Gaussian level.
The results for a two-component model are qualitatively similar, but the details depend on the interactions among the modes.
In these units, the intensity of the anomaly at T$_\mathrm{V}$ is regulated by the inverse Ginzburg length \cite{chaikin2000}.

\newcommand{\coolphi}{_{\,_{\!\!\Large\phi}}}
\renewcommand{\thesection}{S\arabic{section}}  
\renewcommand{\thefigure}{S\arabic{figure}} 
\renewcommand\Im{\operatorname{\mathfrak{Im}}}
\renewcommand\Im{\operatorname{\mathfrak{Im}}}

\DeclareRobustCommand{\rchi}{{\mathpalette\irchi\relax}}
\newcommand{\irchi}[2]{\raisebox{\depth}{$#1\chi$}}

\setlength{\parindent}{0cm}

\clearpage
\pagebreak	
	\begin{center}
		\large\textbf{Coherent generation of symmetry-forbidden phonons by light-induced electron-phonon interactions in magnetite \\ Supplemental Material}
	\end{center}
	
	\section*{S1. SAMPLE CHARACTERIZATION} 
	
\begin{figure}[h]
		\includegraphics[width=0.6\textwidth]{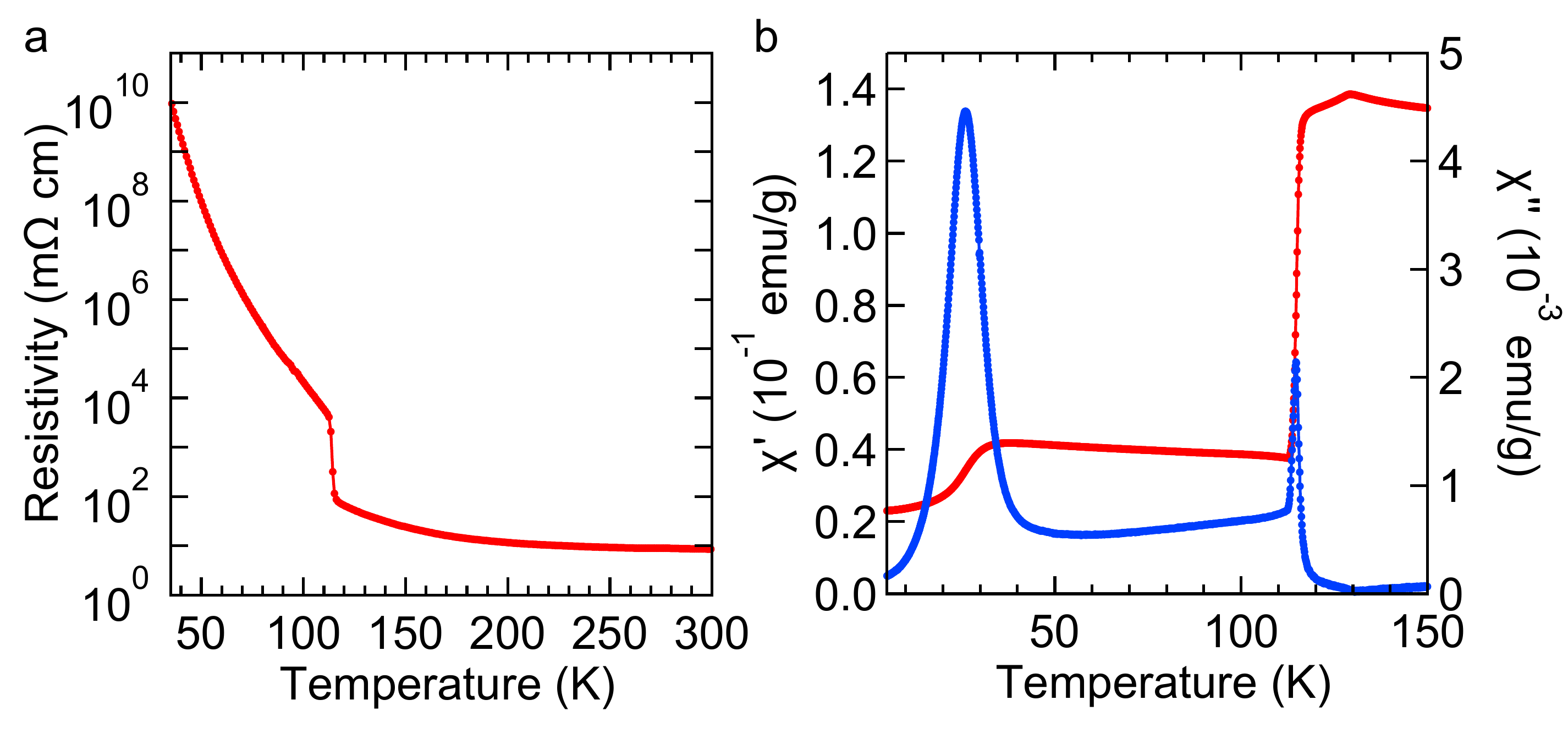}
		\caption{ \label{fig:Resistivity}
			\textbf{Resistivity and ac susceptibility characterization of our sample.} (\textbf{a}) Temperature dependence of the resistivity of our sample. (\textbf{b}) Temperature dependence of the ac susceptibility of our sample (red line –- real part $\rchi$', blue line –- imaginary part $\rchi$''). The frequency and amplitude of the driving field are respectively 7 Hz and 0.64 Oe.
		}
	\end{figure}

	The resistivity and ac magnetic susceptibility of our sample were measured as a function of temperature, respectively in the ranges 35--300 K and 5--150 K, to determine the Verwey temperature (T$_\mathrm{V}$) and the transition order. The ac susceptibility measurements were performed using a commercial CryoBIND system. Resistivity and ac magnetic susceptibility data are displayed in Fig. \ref{fig:Resistivity}. The resistivity data show that the transition is first order and T$_\mathrm{V}$ = 116 K, below typical values for highly stoichiometric samples, owing to the impurity content inherent to natural rocks of magnetite. As illustrated in Fig. \ref{fig:Resistivity}b, the ac magnetic susceptibility of our sample exhibits features analogous to Refs. \citenum{skumryev1999ac} and \citenum{balanda2005magnetic}, referred to synthetic crystals. In particular, in accordance to the above studies, the magnetic response is independent of frequency in the critical region, whereas a pronounced frequency dependence characterizes the susceptibility anomaly at low temperature. The magnetization dynamics is ascribed to domain wall motion (DWM) according to Ref. \citenum{balanda2005magnetic}. The drop in the real part of the ac magnetic susceptibility $\rchi$' originates from crystal microtwinning across the structural transition. In the monoclinic phase, the magnetic domains are coincident with the ferroelastic domains formed below T$_\mathrm{V}$, which require higher fields to propagate via DWM, compared to their cubic counterparts. Therefore, the temperature for the discontinuity in $\rchi$' is understood as the critical temperature of the structural transition, and found to correspond to that of the metal-insulator transition. 
		
	\section*{S2. SPONTANEOUS RAMAN SCATTERING}
	
	Figure \ref{fig:Raman_Spectra} shows the spontaneous Raman spectra of magnetite as a function of temperature at 2.4 and 3.1 eV excitation energies.  
	In the high temperature phase, the T$_\mathrm{2g}$ mode is visible only for 2.4 eV excitation energy, which further substantiates the dependence on energy of its Raman matrix element (RME), as deduced from our time-resolved data.
	The energy of the T$_\mathrm{2g}$ mode is in agreement with Ref 21. 
	The symmetry breaking across the Verwey transition gives rise to a fine structure of Raman-active modes at the Brillouin zone center.
	The red arrows in Fig. S3 mark two peaks which appear below T$_\mathrm{V}$ and are not related to any peak in the high-temperature phase. 
	On the basis of the analysis presented in the main text, the peak at 20.2 meV at 5 K is assigned to the A$_\mathrm{g}$ mode that originates from the $\Delta_2$ mode in the cubic phase. 
	The dependence on temperature of its energy and width is depicted in Fig. S3c. 
	A lower energy mode is barely visible at 15.6 meV at 5 K, only for 2.4 eV excitation energy, owing to  experimental limitations to the lower cut-off energy when using 3.1 eV light.
	Based on the same analysis, this mode is attributed to the lowest-energy X$_3$ mode in the cubic phase.
	
	\section*{S3. AB INITIO CALCULATIONS OF OPTICAL CONSTANTS and RMEs}
	
	The {\sc yambo} code for many-body calculations in solids~[46] was used to compute the optical absorption spectrum within the linear response random phase approximation. 
	The GGA+U ground state (GS) Kohn-Sham states calculated with the {\sc quantum espresso} (QE) suite of programs~[45] were used as input for the optical absorption calculations. The GS self-consistent field (SCF) charge density was obtained using norm-conserving pseudopotentials from the QE-pseudopotential library, a plane wave basis-set with 100 Ry energy cutoff and 560 K-points sampling the first Brillouin zone. 
	The experimentally observed charge ordering on the Fe B-sites was well reproduced with on-site Coulomb repulsion U = 4 eV and Hund's exchange coupling J = 1 eV, yielding an insulating gap of 0.35 eV. 
	As shown in Fig. S5, the optical constants obtained for U = 3.5 eV and U = 5 eV are in poorer agreement with the data compared to those calculated for U = 4 eV.
	
	The RMEs were computed for displacement amplitudes corresponding to once (~$\lambda$~=~1~) and twice (~$\lambda$~=~2~) the mode contributions to the structural distortions from $Fd\bar{3}m$ to $P2/c$ symmetry, starting from the equilibrium positions in the monoclinic structure. The maximum displacements of the atoms for $\lambda$~=~1 are about $8 \times 10^{-3}$ \AA, \: $6 \times 10^{-3}$ \AA \: and $13 \times 10^{-3}$ \AA, respectively for the X$_3$, $\Delta_5$ and $\Delta_2$ modes. The RME calculated for $\lambda$~=~2 and the X$_3$ ($\Delta_2$) mode was multiplied by a factor 3 (2) for the comparison with the experimental RMEs (Fig. 6c--e of the main text), which is equivalent to displacing the atoms by $\lambda$ = 6 (4) if non-linear effects are negligible. Our results for $\lambda$~=~1 and $\lambda$~=~2 show non-negligible non-linear effects, which however remain small in the probed spectral range.
	
	\vfill
	\pagebreak
	\section*{S4. SUPPLEMENTARY FIGURES}
	
	\begin{figure}[h]
		\includegraphics[width=\linewidth]{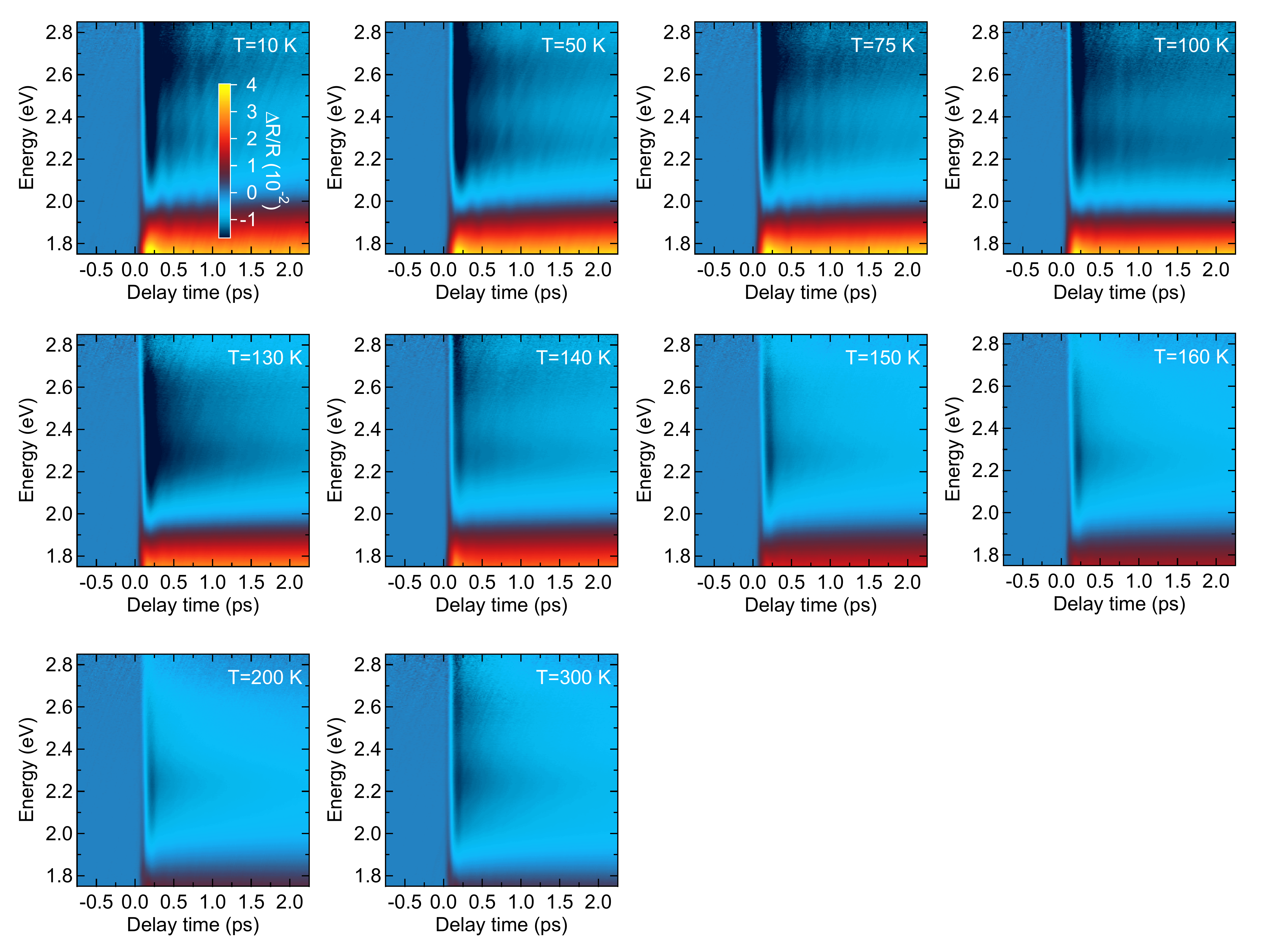}
		\caption{\label{fig:TR-Spectra}
			\textbf{Summary of time-resolved data.} Color-coded maps of differential reflectivity spectra as a function of pump-probe delay time at different temperatures and comparable fluences of $\sim$1 mJ/cm$^2$.}
	\end{figure}
	
	\begin{figure}[h]
		\includegraphics[width=0.8\textwidth]{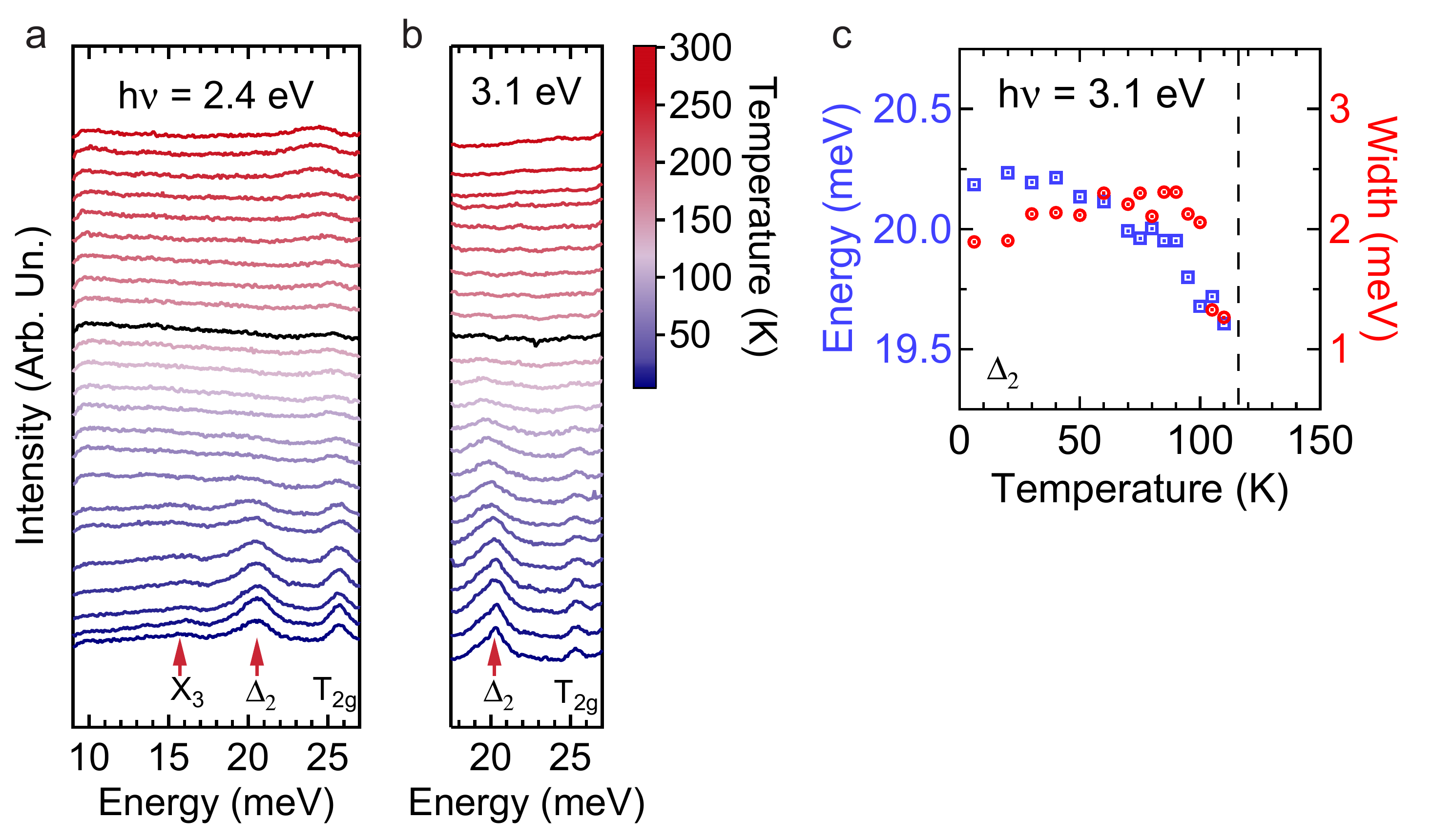}
		\caption{\label{fig:Raman_Spectra}
			\textbf{Summary of spontaneous Raman data.} Spontaneous Raman spectra as a function of temperature at 2.4 eV (\textbf{a}) and 3.1 eV (\textbf{b}) excitation energies. In the color code, black denotes T$_\mathrm{V}$. Red arrows indicate the modes that are nominally optically-active only below T$_\mathrm{V}$. The modes are labeled according to the symmetry of their cubic counterparts. (\textbf{c}) Dependence on temperature of energy and width of the A$_\mathrm{g}$ mode in the monoclinic structure that originates from the folding of the $\Delta_2$ mode of the cubic structure from the $\Delta$- to the $\Gamma$-point. The average width of the mode corresponds to a phonon lifetime of 620$\pm$50 fs.
		}
	\end{figure}
	
	\begin{figure}[h]
		\includegraphics[width=0.6\textwidth]{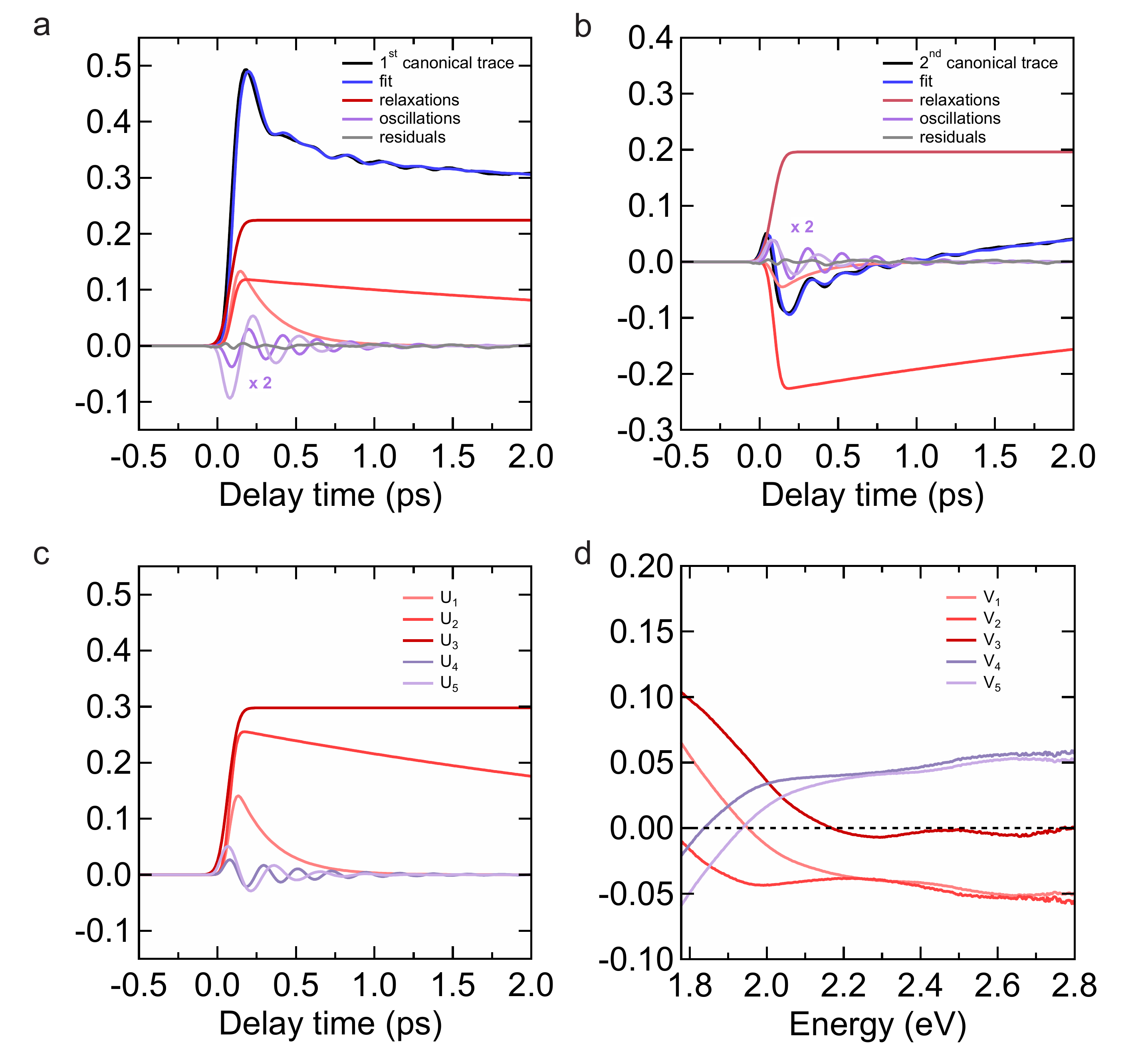}
		\caption{ \label{fig:singular_value_decomposition}
			\textbf{Singular value decomposition (SVD) of the differential reflectivity matrix and fit of the canonical traces with model functions.} Canonical time traces u$_1$(t) (\textbf{a}) and u$_2$(t) (\textbf{b}) obtained from our SVD algorithm. The fitting functions are comprised of the sum of two exponential decays, a step function and two damped coherent oscillations, convolved with a Gaussian response function. Time (\textbf{c}) and energy dependence (\textbf{d}) of the physical traces obtained from SVD, respectively, $\{$U$_i$(t)$\}$ and $\{$V$_i$(E)$\}$, which include relaxations (i=1,2,3) and damped coherent oscillations (i=4,5).
		}
	\end{figure}
	
	\clearpage
	\begin{figure}[h]
		\includegraphics[width=0.45\textwidth]{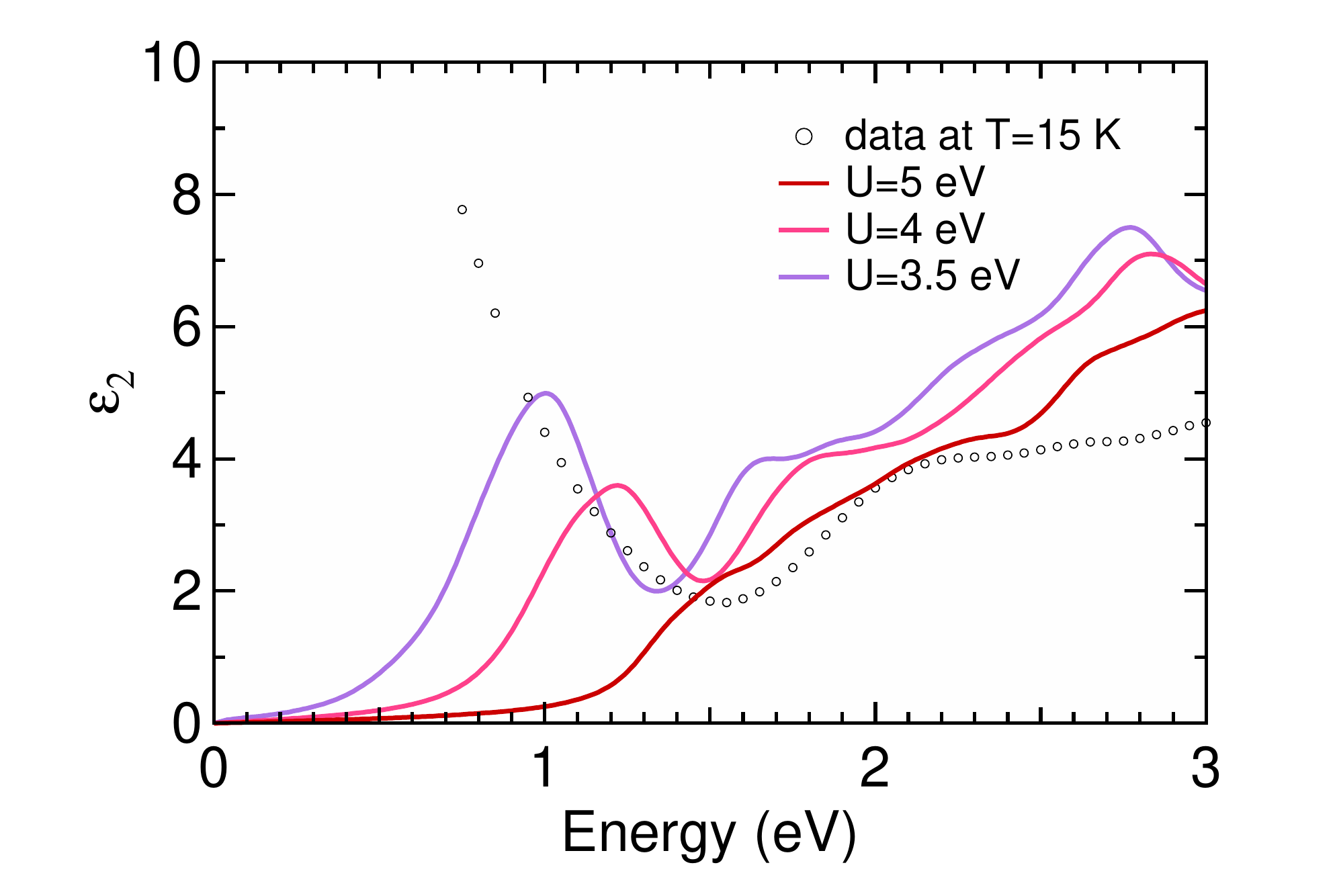}
		\caption{ \label{fig:optical_constants_comparison}
			\textbf{Comparison between data and calculations of the optical response.}
			The imaginary part of the permittivity, $\epsilon_2$, at 15 K from the ellipsometry measurements of Ref. 4 (open circles) is plotted together with the theoretical predictions from our GGA+U calculations for different values of the U parameter and exchange coupling J = 1 eV (lines). 
		}
	\end{figure}
	
	\clearpage
	\begin{figure}[h]
		\includegraphics[width=0.6\textwidth]{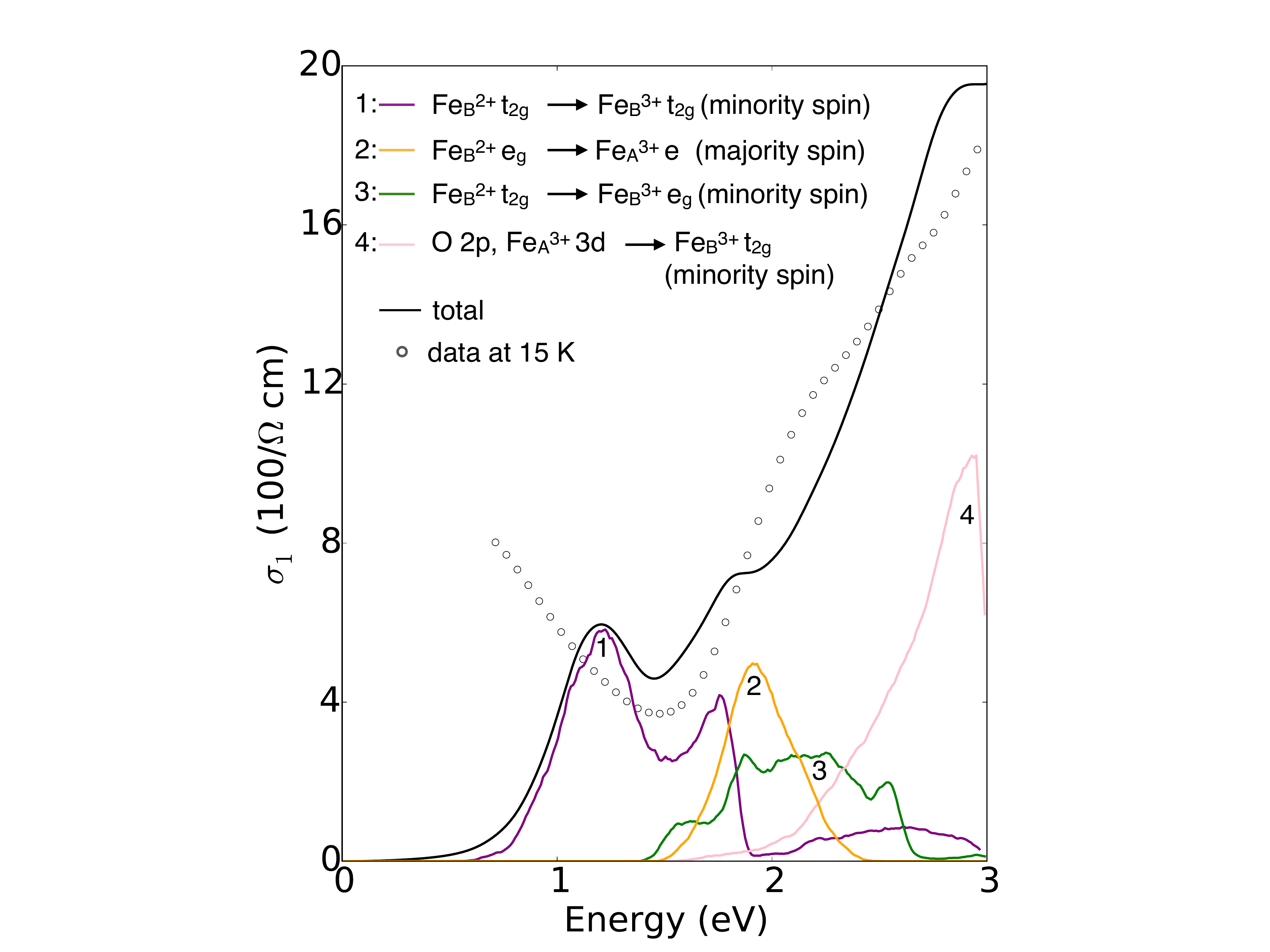}
		\caption{ \label{fig:optical_constants_comparison}
			\textbf{Assignment of the interband transitions below 3 eV.}
			Comparison between the absorptive part of the optical conductivity from experiments (open circles, Ref. 4) and our GGA+U calculations for U = 4 eV and J = 1 eV (black line). The separate contributions from the interband transitions below 3 eV are also plotted (colored lines) and labeled according to the predominant character of the initial and final states.
		}
	\end{figure}
	
	\clearpage
	\begin{figure}[h]
		\includegraphics[width=\textwidth]{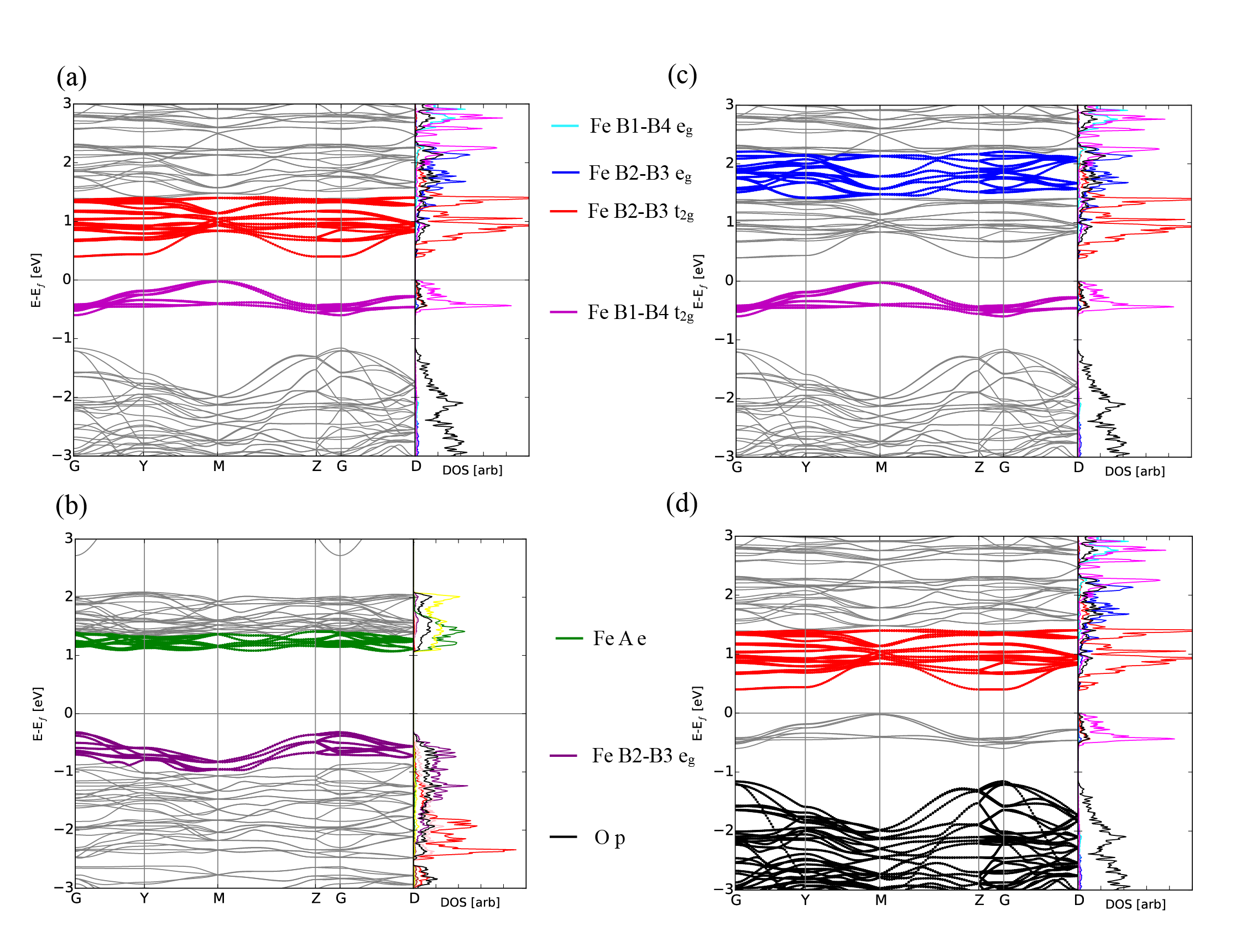}
		\caption{ \label{fig:bands_optical_transitions}
			\textbf{Energy dispersion of the initial and final states of the interband transitions in Fig. S6.} Density of states and band structure of the minority (\textbf{a},\textbf{c},\textbf{d}) and majority spin channel (\textbf{b}) of the low-temperature phase of magnetite in the approximate symmetry $P2/c$ obtained from our GGA+U calculations with U = 4 eV and J = 1 eV. The initial and final states of the interband transitions in Fig. S6 are here highlighted in different colors, namely, peak 1 -- Fe$_B^{2+}$ $t_{2g}$ $\rightarrow$ Fe$_B^{3+}$ $t_{2g}$ (\textbf{a}), peak 2 -- Fe$_B^{2+}$ $e_{g}$ $\rightarrow$ Fe$_A^{3+}$ $e$ (\textbf{b}), peak 3 -- Fe$_B^{2+}$ $t_{2g}$ $\rightarrow$ Fe$_B^{3+}$ $e_{g}$ (\textbf{c}) and peak 4 -- O $2p$ , Fe$_A^{3+}$ $3d$ $\rightarrow$ Fe$_B^{3+}$ $t_{2g}$ (\textbf{d}).
		}
	\end{figure}

\end{document}